\documentclass[11pt,a4paper]{article}
\usepackage{jcappub}

\pdfoutput=1

\usepackage{slashed}
\usepackage{cancel}
\usepackage{textcomp}
\usepackage{calc}

\allowdisplaybreaks

\catcode`\@=11
\def\lsim{\mathrel{\mathpalette\@versim<}}
\def\gsim{\mathrel{\mathpalette\@versim>}}
\def\@versim#1#2{\vcenter{\offinterlineskip
\ialign{$\m@th#1\hfil##\hfil$\crcr#2\crcr\sim\crcr } }}
\catcode`\@=12

\newcommand{\Slash}[1]{{\ooalign{\hfil/\hfil\crcr$#1$}}}

\newcommand{\al}[1]{\begin{align}#1\end{align}}
\newcommand{\bp}{\begin{pmatrix}}
\newcommand{\ep}{\end{pmatrix}}
\newcommand{\nn}{\nonumber}

\newcommand{\bs}[1]{\boldsymbol}

\newcommand{\be}{\begin{eqnarray}}
\newcommand{\ee}{\end{eqnarray}}

\title{Scale Invariant Extension of the Standard Model:\\
A  Nightmare Scenario in Cosmology}

\author[a]{Mayumi Aoki}
\emailAdd{mayumi.aoki@staff.kanazawa-u.ac.jp}
\affiliation[a]{Institute  for  Theoretical  Physics,  Kanazawa  University,  Kanazawa  920-1192,  Japan}

\author[b,c]{Jisuke Kubo}
\emailAdd{kubo@mpi-hd.mpg.de}
\affiliation[b]{Max-Planck-Institut f\"ur Kernphysik (MPIK), Saupfercheckweg 1, 69117 Heidelberg, Germany}
\affiliation[c]{Department of Physics, University of Toyama, 3190 Gofuku, Toyama 930-8555, Japan}

\author[d]{Jinbo Yang}
\emailAdd{yangjinbo@gzhu.edu.cn}
\affiliation[d]{ Department of Astronomy, School of Physics and Materials Science,
	Guangzhou University, Guangzhou 510006, P.R.China}
	
\abstract{
Inflationary observables of  a classically scale invariant model, in which
 the origin of
the Planck mass and the electroweak scale
 including the right-handed neutrino mass is 
  chiral symmetry breaking in a QCD-like hidden sector,
are studied.
Despite a three-field inflation the initial-value-dependence is
strongly suppressed thanks to a river-valley like potential.
The model predicts
the tensor-to-scalar ratio $r$ of cosmological perturbations 
smaller than that of the $R^2$ inflation, i.e.,
$ 0.0044 \gsim r \gsim 0.0017$ for  e-foldings between $50$ and $60$:
The model will be consistent even with a  null detection 
at LiteBird/CMB-S4.  We  find that
the non-Gaussianity parameter $f_{NL}$ 
  is $ O(10^{-2})$,
 the same size  as  that of  single-field inflation.
The dark matter particles are the lightest Nambu-Goldstone bosons
associated with chiral symmetry breaking, which are
   decay products of one of the inflatons and
are heavier than $10^9$  GeV with a strongly suppressed 
 coupling with the standard model,
implying that the dark matter will be 
 unobservable in direct as well as indirect measurements.
 }

\keywords{particle physics - cosmology connection, inflation, non-gaussianity}

\arxivnumber{2401.12442}

\begin{document}
\maketitle

  \newpage
\section{Introduction}
\label{introduction}
 The idea of cosmic inflation \cite{Sato:1980yn,Linde:1981mu,Linde:1982zj,Albrecht:1982wi}
is fully consistent with 
 the Planck and BICEP/Keck data of  CMB  measurements 
 \cite{Aghanim:2018eyx,Planck:2018jri,BICEP:2021}.
 Their results show
that the scalar spectral index $n_s$ of the gravitational fluctuations is close to one, and 
 that the ratio $r$ of tensor to scalar power 
spectra of fluctuations is very small;  $r \lsim 0.035$ at 95\%  confidence level
\cite{BICEP:2021}.
Accordingly, the  $R^2$ inflation-the Starobinsky inflation-\cite{Starobinsky:1980te,Mukhanov:1981xt,Starobinsky:1983zz} 
 and also the Higgs inflation \cite{Bezrukov_2008}
 seem 
to be the most promising candidate models \cite{Planck:2018jri}. 
In fact, future experiments \cite{LiteBIRD:2022cnt,CMB-S4:2020lpa,Achucarro:2022qrl} are planned to arrive at an accuracy of
$O(10^{-3})$ for $r$, where the aforementioned models predict
$r$ to be of the same order.

The main reason for the success of these models is that
their inflationary scalar potential is super-flat, when
expressed in the Einstein frame after a local Weyl scaling 
from the Jordan frame \cite{Bezrukov_2008,Mijic:1986iv} . 
The super flatness in the scalar potential in both models is caused by the relative suppression of the non-scale-invariant Einstein-Hilbert term, $R$,  compared, respectively, to the scale invariant term of the Starobinsky inflation, $\gamma R^2$
with $\gamma \sim O(10^9)$ \cite{Mijic:1986iv,Hwang:2001pu} and to that of the Higgs inflation, $\beta |H|^2 R$ with $\beta \sim O(10^4)$ \cite{Bezrukov_2008}, 
where $R$ is the Ricci curvature 
 scalar and $H$ is the standard model (SM) Higgs doublet. 
 (We note however that a super-flat potential can be realized
for the Higgs inflation even for $\beta \sim O(10)$ if 
the RG improved Higgs potential is used
\cite{Hamada:2014iga,Bezrukov:2014bra}.)

Thus, we may interpret the results of the Planck and BECEP/Keck experiments as suggesting
 a scale invariant extension of Einstein’s theory of gravity.
Along this line of thought a number of models have been recently proposed
 \cite{Salvio:2014soa, Salvio:2020axm, Kannike:2015apa,Farzinnia:2015fka,Karam:2018mft,Kubo:2020fdd,GarciaBellido:2011de,Rinaldi_2016, Ferreira_2016,Ferreira_2019,Tambalo_2017,Benisty:2018fja, Barnaveli:2018dxo, Ghilencea:2018thl, Ghilencea:2019rqj,Kubo:2018kho, Ishida:2019wkd,Kannike:2014mia,Barrie:2016rnv,Vicentini:2019etr,Gialamas:2020snr,Gialamas:2021enw,Aoki:2021skm,Kubo:2022dlx}.
 These models involve a multi-field system for cosmic inflation.
 Therefore, the inflation trajectory that describes the background Universe
is   in a multi-dimensional space of fields, which means that the initial value
of the trajectory is  also multi-dimensional.
Note that there is no warranty a priori that 
the predictions of inflationary parameters
do not largely  depend on the  multi-dimensional   initial values. 
However, to our knowledge,
detailed investigations of  the initial-value-dependence in the models
mentioned   above have not been made.

If the inflationary potential of a multi-field system has 
a  river-valley like structure, there are a slow-roll direction
(the direction along the river) and also  the direction(s) perpendicular to 
it (the heavy-mode direction(s)) \cite{Peterson:2010np,GarciaBellido:2011de},
where slow-roll conditions are likely violated 
in the heavy-mode direction(s).
Furthermore, although the background Universe is described by 
the multi-field system,
we have to take care of the fact that 
the filed fluctuations of the heavy  mode(s) do not produce perturbations on cosmologically relevant scales \cite{Pilo:2004ke,Bartolo:2004if}. 

In this paper, we revisit the classical scale-invariant model of Ref.
\cite{Aoki:2021skm},
in which a strongly interacting QCD-like hidden sector is introduced
and the mass scale is generated in a non-perturbative way via condensation that breaks chiral symmetry dynamically
\cite{Nambu:1960xd,Nambu:1961tp,Nambu:1961fr} in the hidden sector 
\cite{Hur:2011sv,Heikinheimo:2013fta,Holthausen:2013ota,Hatanaka:2016rek,Kubo:2014ida,Ametani:2015jla}.

In Section \ref{Model} we start by briefly  describing the model.
In particular, we review how we use 
 the Nambu-Jona-Lasinio (NJL) model \cite{Nambu:1960xd,Nambu:1961tp,Nambu:1961fr} 
 as an effective low-energy theory in a mean field approximation.
 In Section \ref{inflation} we discuss inflation.
First we derive the effective action for inflation in the Jordan frame.
We next  go to the Einstein frame
and discuss the
river-valley like structure of the inflationary potential
to reduce a three-field system to a two-filed system for inflation.
Then we consider the two-field system by solving 
coupled equations of motion numerically and  analyze 
the initial-value-dependence.
We also identify 
the  light mode (corresponding to the slow-roll direction)
and  the heavy mode to see
how the slow-roll conditions are violated.
Before we proceed with this issue, we briefly outline the basic ingredients of
the  $\delta N$ formalism 
\cite{Sasaki:1995aw,Nakamura:1996da,Sasaki:1998ug,Lyth:2004gb,Lyth:2005fi,Sugiyama:2012tj}.
Then,  adjusting   the $\delta N$ formalism to the aforementioned circumstance,
we  develop an algorithm 
 to evaluate  numerically the derivatives of e-foldings
$N$ with respect to background fields
to compute
inflationary parameters including
 the non-Gaussianity parameter $f_{NL}$ \cite{Komatsu:2001rj,Maldacena:2002vr}
 of the primordial curvature perturbations.

In Section \ref{darkmatter}, we discuss a particular dark matter  (DM)  candidate in the hidden sector, the Nambu-Goldstone (NG) bosons associated with the chiral symmetry breaking, which are
 produced   through the decay of one of the inflatons \cite{Chung:1998rq,Allahverdi:2002nb,Garcia:2020eof}.
The DM  relic abundance depends crucially on the reheating temperature, which
we estimate using the method of Refs. \cite{Liddle:2003as,Martin:2010kz,Martin:2013tda,Planck:2018jri}.
Section \ref{conclusion} is devoted to Conclusion.

\section{The Model} 
\label{Model}
\subsection{Four elements}
\label{four-elements}
The model of Ref. \cite{Aoki:2021skm} that we
 consider  consists of four elements, which  describe, respectively,  (i) the generation 
of a robust energy scale 
through the formation of chiral condensate,
(ii) gravity together with cosmic inflation, and (iii)
the SM interactions which are
 slightly extended because of the Higgs portal coupling, 
 (iv) including the right- and left-handed neutrinos:
\al{
(\mbox{i})~~ &\frac{{\cal L}_H  }{\sqrt{-g}}=	-\frac{1}{2}\mbox{Tr}~F^2+
\mbox{Tr}~\bar{\psi}(i\gamma^\mu  
\slashed{D}_\mu  -\boldsymbol{y} S)\psi \,,
\label{LH}\\
(\mbox{ii})~~ &\frac{{\cal L}_\text{G} }{\sqrt{-g}}  =
-\frac{ \beta}{2}\,S^2\,R +\gamma\,R^2+
\kappa\,W_{\mu\nu\alpha\beta}W^{\mu\nu\alpha\beta} \,,
\label{LCGR}\\
(\mbox{iii})~~ &\frac{{\cal L}_{{\rm SM}+S} }{\sqrt{-g}} = 
\frac{1}{\sqrt{-g}}\left. {\cal L}_{\rm SM}\right|_{\mu_H=0}+\frac{1}{2}g^{\mu\nu}\partial_\mu S \partial_\nu S
+\frac{1}{4}\lambda_{HS} S^2 H^\dag H-\frac{1}{4} \lambda_S S^4\,,
\label{LSM}\\
(\mbox{iv})~~ &\frac{{\cal L}_N  }{\sqrt{-g}}=	 \frac{i}{2} \bar{N}_R \slashed{\partial} N_R 
	- \frac{1}{2} y_M S N^T_R C N_R 	
	-\left( y_\nu \bar{L} \tilde{H} \,
	N_R + \text{h.c.} \right)\,,
\label{LNR}
}
where we have imposed scale invariance, meaning that  no element
 contains any dimensionful parameter at the classical level.
In the Lagrangian (\ref{LH}),
$F$ is the matrix-valued field-strength tensor of the $SU(N_c)_H$ 
gauge theory, 
coupled with the vector-like hidden fermions $\psi_i~(i=1,\dots,n_f)$ belonging to the fundamental representation of  $SU(N_c)_H$, 
\footnote{Due to the presence of the  fermions the use of the vierbein formalism is silently understood.
But it does not play any role in the following discussions.}
 and $S$ is a real SM singlet scalar, where
 we assume $N_c=n_f=3$.
For the part (ii)
we have suppressed the Ricci curvature tensor squared,
$R_{\mu\nu\alpha\beta}R^{\mu\nu\alpha\beta}$, because 
it (and also $R_{\mu\nu}R^{\mu\nu}$)
can be written 
as a linear combination of $R^2$, 
the Weyl tensor squared
$W_{\mu\nu\alpha\beta}W^{\mu\nu\alpha\beta} $ and 
the Gau\ss-Bonnet term which is  a surface term.
\footnote{The Gau\ss-Bonnet term and also the 
surface term $\Box\, R$   
can influence the dynamics of the system
when RG running of coupling 
constants are taken into account \cite{Myrzakulov:2014hca}
(see also \cite{Nojiri:2017ncd}).}
$\left.{\cal L}_{\rm SM}\right|_{\mu_H=0}$ 
in the Lagrangian (\ref{LSM}) stands for the SM Lagrangian with
the Higgs mass term suppressed, where
$H$ is the SM Higgs doublet.
The right-handed neutrino is denoted by 
$N_R$ in the Lagrangian (\ref{LNR}),
where $\tilde{H}=i\sigma_2 H^*$, $L$ is the lepton doublet, and
$C$ is the charge conjugation matrix.

The chiral condensate $\langle \bar{\psi} \psi\rangle$ 
 generates
a linear term in $S$ \cite{Hur:2011sv, Heikinheimo:2013fta, Holthausen:2013ota, Hatanaka:2016rek}, leading to a nonzero 
VEV of $S$ denoted by $v_S$, which
 is responsible for the origin
of the Planck mass as well as  the  right-handed neutrino mass
 $m_N = y_M v_S$.
 \footnote{
Strictly speaking, the Yukawa couplings $y_\nu$ 
and $y_M$ should be matrices in the generation space.  
However, since 
we will not consider the flavor structure,  $y_\nu$ and $y_M$ stand for representative real numbers.}
 We assume that $m_N \sim 10^7$ GeV to obtain a desired size of 
the radiative correction 
to the Higgs mass term
\cite{Vissani:1997ys,Casas:1999cd,Clarke:2015gwa,Bambhaniya:2016rbb}  for triggering the electroweak symmetry breaking
and at the same  time to make the type-I seesaw mechanism 
 \cite{Minkowski,Yanagida:1979as,GellMann:1980vs,Goran} 
viable - a scenario dubbed the “Neutrino option” \cite{Brivio:2017dfq}
(see also \cite{Brivio:2018rzm,Brdar:2018num,Brdar:2018vjq,Brdar:2019iem,Brivio:2019hrj,Aoki:2020mlo,Kubo:2020fdd,Brivio:2020aut}).
Further, the Yukawa coupling in (\ref{LH}) violates explicitly chiral symmetry.
Consequently, the (quasi) NG bosons,
 associated with the dynamical chiral symmetry breaking,
  acquire their masses and can become a DM candidate due 
  to a remnant unbroken vector-like 
 flavor group that can stabilise them \cite{Hur:2011sv, Heikinheimo:2013fta, Holthausen:2013ota, Hatanaka:2016rek}.

In brief summary, the real scalar field $S$, involved in all the  sectors,
first transfers the robust energy scale, created
by  the chiral condensate in the hidden QCD-like sector, to the gravity sector and at the same time
plays   a role of inflaton and produces DM 
trough its decay.
Another channel of the energy scale transfer is the 
right-handed neutrino, which becomes massive through the Yukawa coupling
$S$-$N$-$N$, 
giving rise to
 the electroweak symmetry breaking as well as to the light active neutrino
 mass.

 \subsection{Nambu-Jona-Lasinio description of chiral symmetry breaking}
 \label{NJL}
 In order to  describe
 the dynamical chiral symmetry breaking in the hidden sector,
 \footnote{Linear sigma model is used in Ref. \cite{Hur:2011sv}, and 
 the holographic method is applied in Ref.  \cite{Hatanaka:2016rek}.}
  we use the NJL
theory \cite{Nambu:1960xd,Nambu:1961tp,Nambu:1961fr}.
As announced we employ
$n_f=N_c=3$, because in this case the meson properties 
in hadron physics can be used to reduce the independent parameters
of the NJL theory
\cite{Holthausen:2013ota,Kubo:2014ida,Ametani:2015jla}.
We proceed with our discussion by writing down 
the  NJL Lagrangian for the hidden sector~\footnote{Here we work in the flat space-time.} 
 \cite{Nambu:1960xd,Nambu:1961tp,Nambu:1961fr}:
 \begin{align}
{\cal L}_{\rm NJL}&=\mbox{Tr}~\bar{\psi}(i\gamma^\mu\partial_\mu 
-{\boldsymbol y}S)\psi+2G~\mbox{Tr} ~\Phi^\dag \Phi
+G_D~(\det \Phi+h.c.)\,,
\label{eq:NJL10}
\end{align}
where
$\Phi_{ij}= \bar{\psi}_i(1-\gamma_5)\psi_j=
\frac{1}{2}\sum_{a=0}^{8}
\lambda_{ji}^a\, [\,\bar{\psi}\lambda^a(1-\gamma_5)\psi\,]$,
$\lambda^a (a=1,\dots, 8)$ are the Gell-Mann matrices with
$\lambda^0=\sqrt{2/3}~{\boldsymbol 1}$, and the canonical dimension of
$G\, (G_D)$ is $-$2 (5).
To analyze the theory described by the Lagrangian (\ref{eq:NJL10}),  we use
the self-consistent mean-field (SCMF) approximation
of Refs. \cite{Kunihiro:1983ej,Hatsuda:1994pi}.
To this end, we
define
the mean fields $\sigma_i~(i=1,2,3)$ 
and $\phi_a~(a=0,\dots,8)$  in the ``Bardeen-Cooper-Schrieffer" vacuum as
\begin{align}
\label{varphi}
\sigma_i =- 4 G\left<\bar{\psi}_i \psi_i  \right> ~\mbox{and}~
\phi_a =-2 i G\left<\bar{\psi}_i \gamma_5 \lambda^a\psi_i  \right>\,,
\end{align}
respectively.
Following  Refs. \cite{Kunihiro:1983ej,Hatsuda:1994pi} 
we arrive at  the mean-field Lagrangian $\mathcal{L}_{\text{MFA}}$ in the SCMF approximation:  
  $ \mathcal{L}_{\text{MFA}} =    
   \mathrm{Tr} ~\bar{\psi}(i\Slash{\partial}-M)\psi
   +{\cal L}_\text{int}(\psi,\sigma,\phi_a)$,
where
$\sigma=\sigma_1=\sigma_2=\sigma_3$, 
the constituent fermion mass $M$ is given by 
$M(S,\sigma)= \sigma+yS-\frac{G_D}{8G^2}\sigma^2~
\mbox{with}~y=y_1=y_2=y_3$,
and ${\cal L}_\text{int}(\psi,\sigma,\phi_a)$
can be found in Ref. \cite{Ametani:2015jla}.

The one-loop effective potential can be obtained 
from 
$\mathcal{L}_{\text{MFA}}$ by integrating out the hidden fermions:
\begin{align}
V_{\rm NJL}(S,\sigma)
& = \frac{3}{8G}\sigma^2-
\frac{G_D}{16G^3}\sigma^3
-3N_c I_0(M,\Lambda_H)\,,
\label{eq:Vnjl}
\end{align}
where the  function $I_0$ is given by
\begin{align}
  I_0(M, \Lambda)= \frac{1}{16\pi^2}\left[ \Lambda^4 \ln \left( 1+\frac{M^2}{\Lambda^2 }\right)-M^4 \ln \left( 1+\frac{\Lambda^2 }{M ^2}\right) + \Lambda^2 M ^2\right] \,,
\end{align}
with a four-dimensional momentum cutoff $\Lambda$. 
\footnote{
The one-loop correction $I_0(M,\Lambda)$
 in a curved space time has been calculated in Refs.~\cite{Inagaki:1993ya,Inagaki:1997kz}. 
  It is proportional to
$(\sigma^2/96 \pi^2)\,R$, which is a minimal coupling like
 $\beta\,S^2R$ and can be safely neglected because
 $\beta=O(1) - O(10^4)$ for realistic inflation. }
Note that
the cutoff parameter $\Lambda$ is an additional free physical parameter in the NJL theory.
We obtain non-zero VEVs, i.e., $v_\sigma=\left<\sigma\right>\ne 0$
and $v_S=\left<S \right>\ne 0$,
for a certain interval of  the dimensionless 
parameters $G^{1/2}\Lambda$ and $(-G_D)^{1/5}\Lambda$
\cite{Holthausen:2013ota,Kubo:2014ida,Ametani:2015jla}:
The non-zero $v_\sigma$ means 
the chiral symmetry breaking in the NJL theory.

The actual value  of $\Lambda$ can be fixed, once the hidden sector is connected
with a sector whose energy scale is known.
In our case, the hidden sector is coupled via the mediator $S$ 
with the gravity sector (ii) described by the Lagrangian
(\ref{LCGR}) as well as  with the right-handed neutrino sector (iv) 
described by the Lagrangian (\ref{LNR}),
while the coupling with the SM sector (iii)
 is assumed to be extremely suppressed,
because we assume that the portal coupling $\lambda_{HS}$ is negligibly small.
The cutoff  in the hidden sector  $\Lambda_H$ can be fixed in the following way.

The NJL parameters for the SM  hadrons satisfy
the dimensionless relations
\cite{Holthausen:2013ota,Kubo:2014ida,Ametani:2015jla}
$ \left.G^{1/2}\Lambda\,\right|_\text{Hadron}=1.82~\mbox{and}~
\left.(-G_D)^{1/5}\Lambda\,\right|_\text{Hadron}=2.29$\,,
where $\Lambda_\text{Hadron} \simeq 1$ GeV.
We assume that  
 the above dimensionless relations remain satisfied
 while scaling-up the values of
$G$, $G_D$ and the cutoff $\Lambda$ from QCD hadron physics.
That is, we assume that
\begin{equation}
 \left. G^{1/2}\Lambda  \,\right|_\text{Hidden} =1.82
 ~~\mbox{and}~~\left. (-G_D)^{1/5}\Lambda\,\right|_\text{Hidden}=2.29
 \label{NJL para}
\end{equation}
are valid even if  $\Lambda_\text{Hidden}=\Lambda_H$
 is much larger than 
$\Lambda_\text{Hadron}$.  
Therefore, the free parameters
$G, G_D$ are fixed when  
the hidden sector scale $ \Lambda_H$ is fixed.
Consequently, the only free parameter in the hidden sector
is the Yukawa coupling $y$ once  $ \Lambda_H$ is fixed.
Since the (reduced) Planck mass becomes
$M_\text{Pl}\simeq \sqrt{\beta}v_S$ and we can  write
$v_S=
c_S(y)\, \Lambda_H$, while the dimensionless quantity $c_S$ is  calculable
for a given $y$, 
we can always  relate $\Lambda_H$ to $M_\text{Pl}$.

We recall that the cutoff scale $\Lambda_H$  is  the energy scale
around which   the hidden QCD-like sector starts to be strongly interacting
so that we may use the NJL theory as a low-energy effective theory
below that scale.
We note however that by this  energy scale   a dynamical energy
is meant,
because  static energy like the zero point energy of a potential
does not influence QCD-like interactions.
Since dynamical energies are created after the end of inflation,
we should require that the scale of these energies be less than $\Lambda_H$.
Further, the spontaneously generated Planck scale $M_\text{Pl}$ is not a dynamical scale,
but the VEVs, $v_\sigma$ and $v_S$, should be less than $\Lambda_H$.
We also  note that it is not only the energy scale,
which plays the role for the applicability of the NJL theory, but also
the scale of the explicit chiral symmetry breaking, because
the NJL theory is an effective theory for chiral symmetry breaking.
In our case it is the Yukawa coupling $y$,
while in QCD   the current quark masses are responsible.
Therefore, since $m_s(\mbox{strange quark mass})/\Lambda_\text{Hadron}
\simeq 0.09\, \mbox{GeV}/
1 \,\mbox{GeV}\simeq 0.1$, the self-consistency requires  $y S/\Lambda_H \lsim 0.1$
during inflation, which we shall check when discussing inflation.

We emphasize that the mean fields $\sigma$ and $\phi_a$ are non-propagating classical fields at the tree level.
Their kinetic terms are generated by integrating out the hidden fermions at the one-loop level.
Furthermore, one can see that 
the potential $V_{\rm NJL}(S,\sigma)$ is asymmetric in $\sigma$, 
which is the reason that the chiral phase transition can become of first-order 
in the NJL theory \cite{Aoki:2017aws,Helmboldt:2019pan,Aoki:2019mlt}.

 \subsection{Planck mass}
 In this subsection we discuss how 
the Planck mass is generated.
We may start by integrating out the quantum fluctuations $\delta S$  at one-loop 
to obtain the effective potential
  \al{
  	U_S(S,R)&=   \frac{1}{4}\lambda_S S^4
  	+\frac{1}{64 \pi^2}\left(\,
  	\tilde{m}_s^4 \ln  [\tilde{m}_s^2/\mu^2]
  	\,\right)\,,
  	\label{Ueff}}
  where $\tilde{m}_s^2 =
  	3 \lambda_S S^2+\beta R$,
  and we have employed the $\overline{\mbox{MS}}$ scheme, while
  the constant $-3/2$ is absorbed into the renormalization scale $\mu$.
  \footnote{
  Strictly speaking,  $\beta$ in $\tilde{m}_s^2$  should read $\beta-1/6$, 
  if one properly takes into account the non-flatness of the space-time background
   \cite{Markkanen:2018bfx}. 
  However, since $\beta$ should be large (i.e., $\gsim 10$) 
  for realistic cosmic inflation \cite{Aoki:2021skm}, we will be ignoring the constant $1/6$ throughout the paper.
  }
So, the total effective potential in the Jordan frame is given by
$U_\mathrm{eff}(S,\sigma,R) = 
V_\text{NJL}(S,\sigma) +U_S(S,R)-U_0$,
where $V_\text{NJL}$ is given in Eq.~(\ref{eq:Vnjl}), and $U_0$ is the zero-point energy density.
We have subtracted it, such that
$U_\mathrm{eff}(S=v_S, \sigma=v_\sigma, R=0)=0$ is satisfied.
Note that the zero-point energy density $U_0$
 is  the cosmological constant at the tree-level
(which breaks  super-softly  the scale invariance).
In other words, we put  the cosmological constant problem
\cite{Weinberg:1988cp}  aside and ignore the existence of the problem.

  To compute  $v_S=\langle S\rangle$ and 
   $v_\sigma=\langle \sigma\rangle$, 
   we assume  that $\beta  R  < 3 \lambda_S S^2 $ (during inflation), such that
$U_S(S,R)$ in  Eq.~(\ref{Ueff}) can be expanded 
  in powers of 
  $\beta R$:
  \al{
  	U_S(S,R) &=
  	U_\mathrm{CW}(S)+ U_{(1)}(S) \,R+U_{(2)}(S) R^2 +O(R^3 )\,,
  	\label{Ueff1}
  }
  where 
  \al{  
  	U_\mathrm{CW}(S) &=\frac{1}{4}\lambda_S S^4+
	\frac{9}{64 \pi^2}
  	 \, \lambda_S^2 S^4\,\left(\,
	 \ln [\,3 \lambda_S S^2/v_S^2\,]-\frac{1}{2}\,\right)\,,\label{Ucw}\\
  	U_{(1)}(S) &=\frac{3}{32 \pi^2}
  	\beta  \lambda_S S^2\,\ln [\,3 \lambda_S S^2/v_S^2\,]\,,
  	\label{U1}\\
  	U_{(2)}(S) &= \frac{1}{64 \pi^2}
  \beta^2  \left(1+\ln [\,3 \lambda_S S^2/v_S^2\,]\right)\,.
  	\label{U2}
  } 
We have chosen $\mu^2=v_S^2/2$,  because for this choice
the logarithmic term in (\ref{Ucw}) does not enter into the  determination of $v_S$
and $v_\sigma$.
 Finally,  the  identification of $M_\mathrm{Pl}$ follows from 
  the first term in Eq.~(\ref{LCGR}) along with Eq.~(\ref{Ueff1}): 
  \al{
  	M_\mathrm{Pl} &= v_S \left( \beta +
  	\frac{2 U_{(1)}(v_S)}{v_S^2} \right)^{1/2} =
	\sqrt{\beta}\, v_S\left(
 1+\frac{3\lambda_S}{16\pi^2}\ln[3\lambda_S]\right)^{1/2}\,.
  	\label{mpl}
  }
Since  $\beta$ is larger than $O(10^2)$ for a successful inflation,
 $v_S$ is at most $\sim M_\mathrm{Pl}/10$.

\section{Inflation}\label{inflation}
\subsection{Effective action for inflation}

If the inequality $\beta  R  < 3 \lambda_S S^2 $ is satisfied during inflation,
the higher order terms in Eq.~(\ref{Ueff1}) 
can be  consistently neglected for inflation. We will proceed with this 
simplification.
Similarly, if $\kappa$, the coefficient
of the $W_{\mu\nu\alpha\beta}W^{\mu\nu\alpha\beta}$ term
in the Lagrangian (\ref{LCGR}), is small, this term  has only a small effect on the 
inflationary parameters (see for instance Refs. \cite{Baumann:2015xxa,Salvio:2017xul,Ghilencea:2019rqj,Anselmi:2020lpp,Salvio:2020axm,Gialamas:2020snr,Kubo:2022dlx} and also \cite{DeFelice:2023psw}),
so that we will ignore it in the following discussion.
Hence
the effective Lagrangian  for inflation in the Jordan frame can be written as
\al{
	\frac{{\cal L}_\mathrm{eff}}{\sqrt{-g_J}} 
	=-\frac{1}{2}M_\mathrm{Pl}^2 B(S) R_J+
	G(S) R_J^2+\frac{1}{2}g_J^{\mu\nu}\left(\partial_\mu S\partial_\nu S+
	Z_\sigma^{-1}(S,\sigma) \partial_\mu \sigma\partial_\nu \sigma
\right)
	-U(S,\sigma)\,,
	\label{Leff}
}
where the subscript $J$ means quantities in  the Jordan-frame, and
\al{
  B(S) =&
\frac{\beta\, S^2}{M_\text{Pl}^2}+
 \frac{2}{M_\text{Pl}^2}U_{(1)}(S)\,,~
  G(S) =\gamma
-U_{(2)}(S)\,,
 \label{GoS}\\
 U(S,\sigma) =&V_\text{NJL}(S,\sigma)+U_\text{CW}(S)-U_0\,.\label{UoS}
 }
To go the Einstein frame we first rewrite the $R_J^2$ term in the Lagrangian (\ref{Leff})
by
introducing an auxiliary field $\chi$ with mass dimension two
and by replacing 
$ G(S) R_J^2$ by 
$2  G(S) R_J\chi-G(S) \chi^2$.
Then we perform a Weyl rescaling of the metric,
$g_{\mu\nu} = \Omega^2\, g_{\mu\nu}^J$ with
$\Omega^2(S,\chi) =
B(S) - 4\,G(S)\chi/M_{\rm Pl}^2$.
Using the scalaron field $\varphi$~\cite{Barrow:1988xh,Maeda:1988ab}, 
which is canonically normalized and  defined as
$\varphi = \sqrt{\frac{3}{2}}\,M_{\rm Pl} \ln\left|\Omega^2\right|$,
we finally obtain the Einstein-frame Lagrangian for the coupled $S$-$\sigma$-scalaron system:
\begin{align}
\label{eq:LEphichi}
\frac{\mathcal{L}_{\rm eff}^E}{\sqrt{- g}} =& -\frac{1}{2}\,M_{\rm Pl}^2\,R
+ \frac{1}{2}\,g^{\mu\nu}\,\partial_\mu\varphi\,\partial_\nu \varphi 
+ \frac{1}{2}\,e^{-\Phi(\varphi)}\,g^{\mu\nu}\left(\,\partial_\mu
S\,\partial_\nu S+Z_\sigma^{-1}(S,\sigma)\partial_\mu \sigma\,\partial_\nu \sigma
\right)\nn\\
& - V(S,\sigma,\varphi)  \,,
\end{align}
where \footnote{$Z_\sigma$ is the wave-function renormalization
constant for $\sigma$, which is computed e.g., in Ref.~\cite{Helmboldt:2019pan}.
In the following sections, however, we will ignore it,
because we will be using the approximation
that $\sigma$ remains unchanged, i.e., fixed at $v_\sigma$, during inflation.
} 
\begin{align}
V(S,\sigma,\varphi) = 
e^{-2\,\Phi(\varphi)} \left[ U(S,\sigma)
+ \frac{M_{\rm Pl}^4}{16\,G(S)}\left(B(S)
- e^{\Phi(\varphi)}\right)^2\right] \,,
\label{VSphi}
\end{align}
with $\Phi\left(\varphi\right) = \sqrt{2/3}\,\varphi/M_{\rm Pl}$.
The form of the potential (\ref{VSphi}) is quite general:
One obtains it, after going to the Einstein frame
if a $R^2$ term is included in the Lagrangian and is converted 
into a term linear in $R$ by introducing an auxiliary field  in the Jordan frame.
The auxiliary field becomes a dynamical  degrees of freedom,
the scalaron \cite{Starobinsky:1980te},  in the Einstein frame.
(See e.g. the scale invariant models of 
\cite{Salvio:2014soa, Salvio:2020axm, Kannike:2015apa,Farzinnia:2015fka,Karam:2018mft,Kubo:2020fdd,
Ferreira_2019,Tambalo_2017,Barnaveli:2018dxo,Ghilencea:2019rqj,Kubo:2018kho,
Vicentini:2019etr,Gialamas:2020snr,Gialamas:2021enw,Aoki:2021skm,Kubo:2022dlx}.)
Note however that the definition of the scalaron is not 
always the same, where we adapt the definition
of Refs. \cite{Barrow:1988xh,Maeda:1988ab}.
The main difference of our potential $U(S,\sigma)$ from
those of the other scale invariant models is that $U(S,\sigma)$ contains a linear term in $S$,
which originates from the NJL part $V_\text{NJL}(S,\sigma)$.
This will be the main reason that our model can predict a tensor-to-scalar ratio
$r$ which is smaller than that of the $R^2$ infaltion, 
as we will see in \ref{result}.

\subsection{River-valley like structure of the potential}
As we see from the effective Lagrangian (\ref{eq:LEphichi}) 
we have a multi-field system at hand \cite{Wands:2007bd}.
It has been found in Ref. \cite{Aoki:2021skm} 
that the potential (\ref{VSphi}) has a river-valley like structure, such that 
the valley approximation \cite{Kannike:2015apa,Kubo:2018kho,Kubo:2020fdd}
can be successfully applied.
In this approximation the multi-field system can be reduced
to a single-field system, which we have analyzed  in 
Ref.~\cite{Aoki:2021skm}. 

\begin{figure}[ht]
\begin{center}
\hspace{0.7cm}
\includegraphics[width=2.5in]{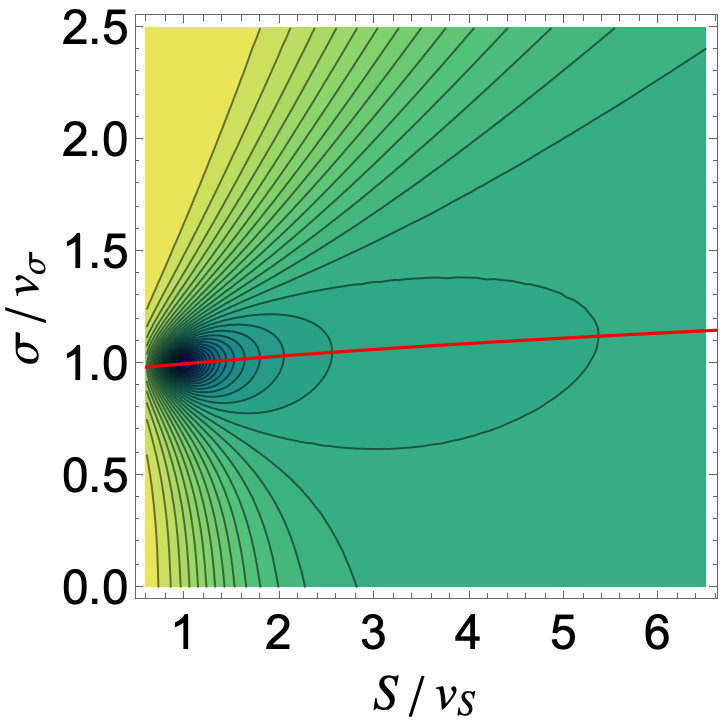}\hspace{0.9cm}
\caption{The contour plot of $\tilde{V}(S,\sigma)$, where
$\tilde{V}(S,\sigma)=V(S,\sigma,\varphi_\text{v}(S,\sigma))$. The red line is
the bottom line of   $\tilde{V}(S,\sigma)$.
}
\label{Vtilde-contour}
\end{center}
\end{figure}

To see the river-valley like structure, we solve
 the stationary point condition\\
$\left.\partial V(S,\sigma,\varphi)/\partial \varphi\,\right|_{ \varphi= \varphi_\text{v}} =0$
with respect to $\varphi$ to obtain
\al{
\varphi_\text{v}(S,\sigma)=&\sqrt{3/2}M_\text{Pl}\, \ln
[\,B(S)+4 A(S,\sigma) B(S)\,]\,,~\mbox{where}~ A(S,\sigma)=\frac{4 G(S) U(S,\sigma)}{B^2(S)M_\text{Pl}^4}\,.
\label{phiv}
}
We then consider the two-field system potential 
$\tilde{V}(S,\sigma)  =V(S,\sigma,\varphi_\text{v}(S,\sigma))$.
In Fig.~\ref{Vtilde-contour}  we show a contour plot of $\tilde{V}(S,\sigma)$ 
for
\al{
y=4.00\times 10^{-3}\,,\quad \lambda_S=1.14\times 10^{-2}\,,\quad
\beta =6.31\times 10^3\,,\quad \gamma =1.26\times 10^8\,.
\label{bench1}
}
The red line in Fig.~\ref{Vtilde-contour}  is the bottom line of 
$\tilde{V}(S,\sigma)$, 
from  which we see that $\sigma$ does not change very much 
during inflation.
 We will use indeed this fact in the following sections to 
 reduce the three-field system to a two-field system:
 We will assume that  $\sigma$ stays at $v_\sigma$ during inflation.

\begin{figure}[ht]
\begin{center}
\hspace{0.7cm}
\includegraphics[width=2.6in]{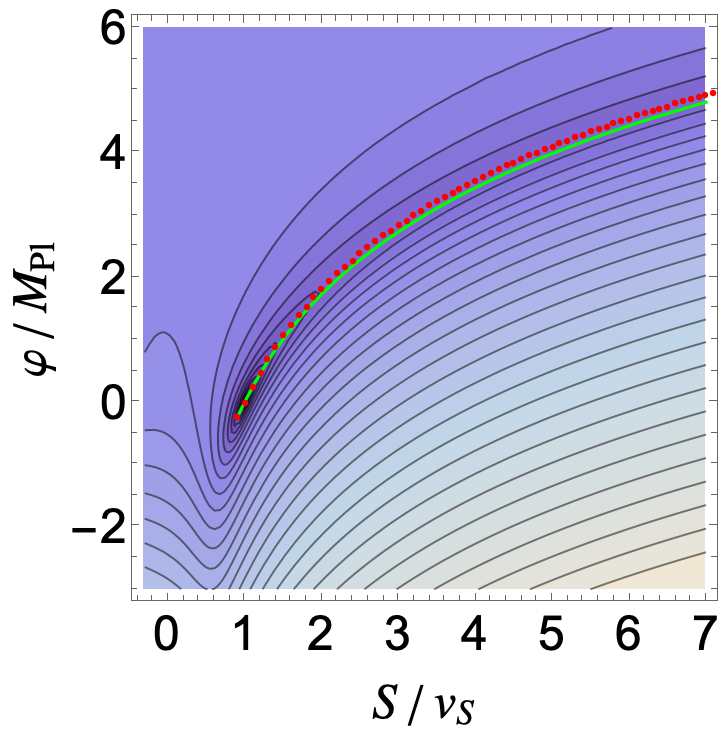}\hspace{0.5cm}
\includegraphics[width=2.7in]{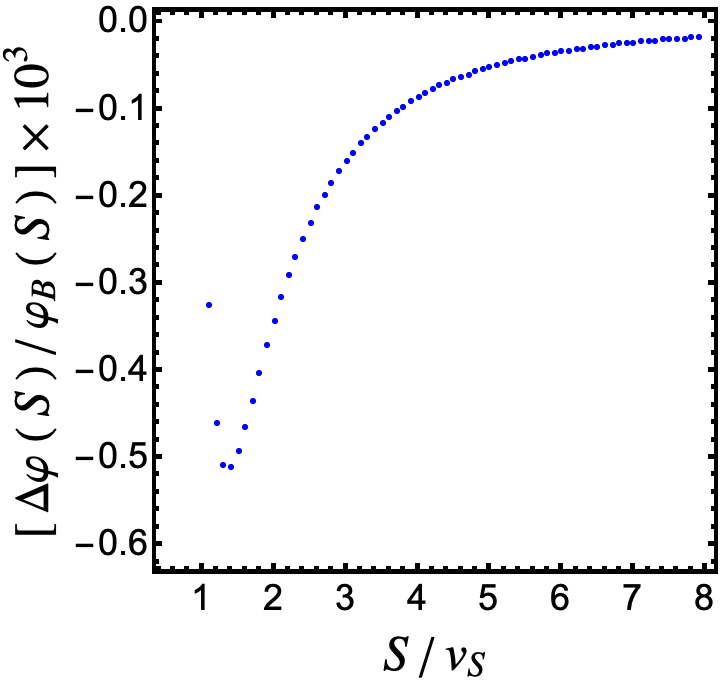}
\caption{Left: The contour plot of $V(S,\sigma_\text{v}(S),\varphi)$,
where $\sigma_\text{v}(S) $ is the bottom line of  $\tilde{V}(S,\sigma)$
(the red line of Fig.~\ref{Vtilde-contour} ),
 where of  $V(S,\sigma,\varphi)$ 
is given in Eq.~(\ref{VSphi}).
The green line is $\varphi_\text{v}(S)$ given in Eq.~(\ref{phiv}), while the dotted red line
(denoted by $\varphi_B(S)$)
is the true bottom line of
 $V(S,\sigma_\text{v}(S),\varphi)$. 
Right: $\Delta \varphi (S)/\varphi_B(S)=
(\varphi_B(S)-\varphi'_B(S))/\varphi_B(S)$ against $S/v_S$,
where  $\varphi'_B(S)$ is the bottom line of  $V(S,\sigma=v_\sigma,\varphi)$.
}
\label{V-S-phi}
\end{center}
\end{figure}

We next consider the potential $V(S,\sigma_\text{v}(S),\varphi)$,
where $\sigma_\text{v}(S)$
is the bottom line of $\tilde{V}(S,\sigma)$, i.e.,
the red line in Fig. \ref{Vtilde-contour} .
Its contour plot is shown in Fig. \ref{V-S-phi} (left) for the same set of the parameters 
(\ref{bench1}).
The green line is $\varphi_\text{v}(S,\sigma_\text{v}(S))$, i.e., (\ref{phiv})
with $\sigma$ replaced by $\sigma_\text{v}(S)$, while the dotted red line 
presents the true bottom line of
 $V(S,\sigma_\text{v}(S),\varphi)$. 
 In the valley approximation in Ref.~\cite{Aoki:2021skm} it is assumed that 
 inflaton rolls down along the green line (see also Ref. \cite{Kubo:2020fdd}).

To figure out the difference between  $V(S,\sigma_\text{v}(S),\varphi)$
and  $V(S,\sigma=v_\sigma,\varphi)$, we consider the bottom line
of these potentials.
We denote the bottom line of  $V(S,\sigma_\text{v}(S),\varphi)$
(the dotted red line in Fig.~\ref{V-S-phi} (left)) 
by $\varphi_B(S)$ and that of 
 $V(S,\sigma=v_\sigma,\varphi)$ by $\varphi'_B(S)$.
  We plot the difference $\Delta \varphi (S)/\varphi_B(S)=
(\varphi_B(S)-\varphi'_B(S))/\varphi_B(S)$ in Fig.~\ref{V-S-phi} (right),
from which we see that the difference is very small, i.e.,
$|\Delta \varphi (S)|/ \varphi_B (S)\sim O(10^{-4})$.
This  also justifies the assumption
that $\sigma$ stays constant fixed at $v_\sigma$ during inflation.


\subsection{Calculation of inflationary observables }
\subsubsection{Background system}
The starting Lagrangian, that describes  our flat and homogeneous background Universe in the Einstein frame,
 is  (\ref{eq:LEphichi}), and  we use it to compute inflationary observables
 including 
the non-Gaussianity of the curvature perturbation 
\cite{Bartolo:2004if} (see also \cite{Baumann:2009ds}).
The original system contains three fields, $S\,,\sigma$ and $\varphi$.
As we see from Fig.~\ref{Vtilde-contour}, $\sigma$ stays approximately constant during
 inflation. Therefore, we set $\sigma$ equal to $v_\sigma$
 to simplify the numerical integration
 of the equations of motion, where $v_\sigma=\langle \sigma\rangle$ 
 can be computed from the potential (\ref{UoS})  
 together with $v_S=\langle S\rangle$.
Then we rewrite  the Lagrangian (\ref{eq:LEphichi})
(with $\sigma$ replaced by $v_\sigma$)
in a geometric form using  the metric ${\cal G}_{IJ}$ 
in the space of the fields:
\begin{equation}
\frac{{\cal L}_\text{eff}^E}{\sqrt{-g}}
=-\frac{1}{2}M_\text{Pl}^2\,R+\frac{1}{2}{\cal G}_{IJ}
g^{\mu\nu}\,\partial_\mu \phi^I \partial_\nu \phi^J
-V(\phi^I)\,,\label{eq:LEG}
\end{equation}
where the ``coordinates'' are 
$\phi^I=(\varphi\,, S)$, and the non-vanishing elements 
of  ${\cal G}_{IJ}$ are
\begin{equation}
{\cal G}_{\varphi\varphi}=1\,,{\cal G}_{SS}=e^{-\Phi(\varphi)}~~
\mbox{with}~~\Phi(\varphi)=\sqrt{2/3} \varphi/M_\text{Pl}\,.
\end{equation}
The equations of motion, that follow from (\ref{eq:LEG}),
are:
\begin{equation}
\ddot{\phi}^I+\Gamma^I_{JK}\,\dot{\phi}^J\dot{\phi}^K+
3 H \,\dot{\phi}^I+{\cal G}^{IJ}\,V_{,J}
=0\label{EQM3}
\end{equation}
with the non-vanishing components of 
$\Gamma_{JK}^I$:
\begin{equation}
\Gamma_{SS}^\varphi=\frac{1}
{\sqrt{6}M_\text{Pl}}e^{-\Phi(\varphi)}\,,~~~
\Gamma_{\varphi S}^S=\Gamma_{S\varphi }^S=\frac{-1}
{\sqrt{6}M_\text{Pl}}\,.
\end{equation}
In terms of the component fields the equations of motion (\ref{EQM3})
become
\begin{eqnarray}
& & \ddot{\varphi} +3 H \dot{\varphi}+
\big(1/\sqrt{6}M_\text{Pl}\big)\,e^{-\Phi(\varphi)}\dot{S}^2 
+V_{,\varphi}=0\label{eq-motion-phi}\,,\\
& &\ddot{S} +3 H \dot{S}
-2\big(1/\sqrt{6}M_\text{Pl}\big)\,\dot{\varphi}\dot{S} +V_{,S}=0\,,
\label{eq-motion-S}
\end{eqnarray}
 where $H$ is the Hubble parameter.
In the subsequent discussions we assume that the fields
are homogenous and depend only on time $t$.
Under this assumption the Hubble parameter $H$ satisfies
\begin{equation}
H^2=\left(\frac{\dot{a}}{a}\right)^2=\frac{1}{3M_\text{Pl}}\Big(
\frac{1}{2}{\cal G}_{IJ}\dot{\phi}^I \dot{\phi}^J
+V(\phi^I)\Big)
\label{Hubble3}
\end{equation}
in the Friedmann-Lema\^itre-Robertson-Walker metric, i.e.,
$ds^2=dt^2-a^2(t) \,d{\bf x}^2$.
Similar models have been considered for instance in Refs.~\cite{Mori:2017caa,Karciauskas:2018urs}
to calculate the non-Gaussianity of the curvature perturbation.
The crucial difference compared with  Refs.~\cite{Mori:2017caa,Karciauskas:2018urs}
is that our potential $V(\phi^I)$ in 
(\ref{eq:LEG}) boasts a river-valley like structure, which causes
complications in calculating the inflationary parameters
without using the valley approximation,  as we will see.

We solve Eqs.~(\ref{eq-motion-phi}) and (\ref{eq-motion-S}) numerically
with a set of initial values of $\varphi$ and $S$ at $t=t_0$.
The end of inflation is defined
such that  $\epsilon (t_\text{end})=1$, where the slow-roll parameter
$\epsilon (t)$
is defined as
\begin{equation}
\epsilon (t)=
\frac{{\cal G}_{IJ}\dot{\phi}^I(t) \dot{\phi}^J(t)}{2M^2_\text{Pl}H^2(t)}
=\frac{1}{2M^2_\text{Pl}H^2(t)}\Big(
\dot{\varphi}(t) \dot{\varphi}(t)+
e^{-\Phi(\varphi(t))}\dot{S}(t) \dot{S}(t)\Big)
\label{epsilon3}\,,
\end{equation}
and the number of e-folds $N$ 
can be calculated from
\begin{equation}
N(t)=
\int^{t_\text{end}}_{t}H(t')\,dt'\,.
\label{e-Fold3}
\end{equation}
During the numerical integration of
Eqs.~(\ref{eq-motion-phi}) and (\ref{eq-motion-S}) 
we check whether the slow-roll conditions 
$\epsilon(t) \ll 1$ and $\beta_\text{sr}(t) \ll 1$
for $t < t_\text{end}$ are satisfied (except for $t$ close to
$t_\text{end}$), where \cite{Nakamura:1996da}
\begin{equation}
\beta_\text{sr}(t)=\frac{2{\cal G}_{IJ}\dot{\phi}^I (\ddot{\phi}^J+
\Gamma^I_{JL}\dot{\phi}^J
\dot{\phi}^L)}{{\cal G}_{KM}\dot{\phi}^K\dot{\phi}^M\, H}\,.
\label{beta-sr}
\end{equation}
(The another slow-roll parameter $\eta$ can be obtained from
$\eta=2\epsilon+2\beta_\text{sr}$.)

\subsubsection{Exact solution vs valley approximation}
\begin{figure}[t]
\begin{center}
\includegraphics[width=7cm]{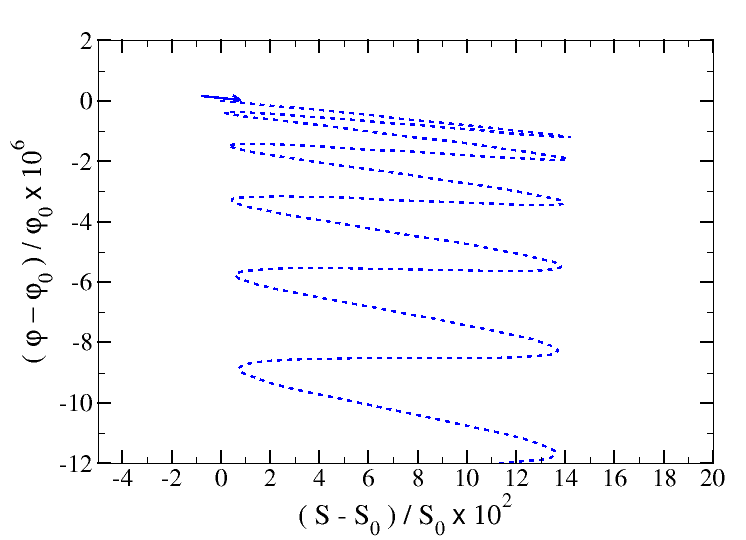}
\includegraphics[width=7cm]{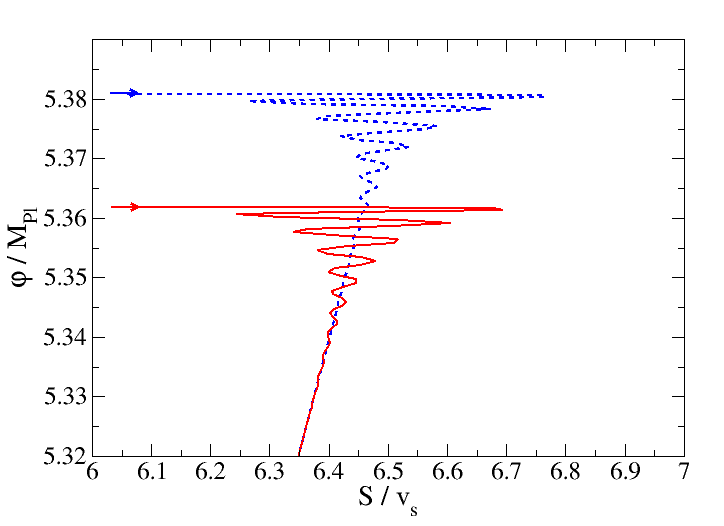}
\caption{Left: Trajectory in the $S$-$\varphi$ plane
just after the start at $t=t_0$. The initial values are given in 
(\ref{in-Value}). The trajectory is strongly oscillating.
Right: Trajectories with two different initial values, 
where the blue dashed  line is for (\ref{in-Value}) 
and the red one is for $S_0/v_S=6.06\, ,\varphi_0/M_\text{Pl}=5.36$.
They converge very fast to a ``fixed-point'' trajectory.}
\label{s-phi}
\end{center}
\end{figure}

\begin{figure}[ht]
\begin{center}
\includegraphics[width=8cm]{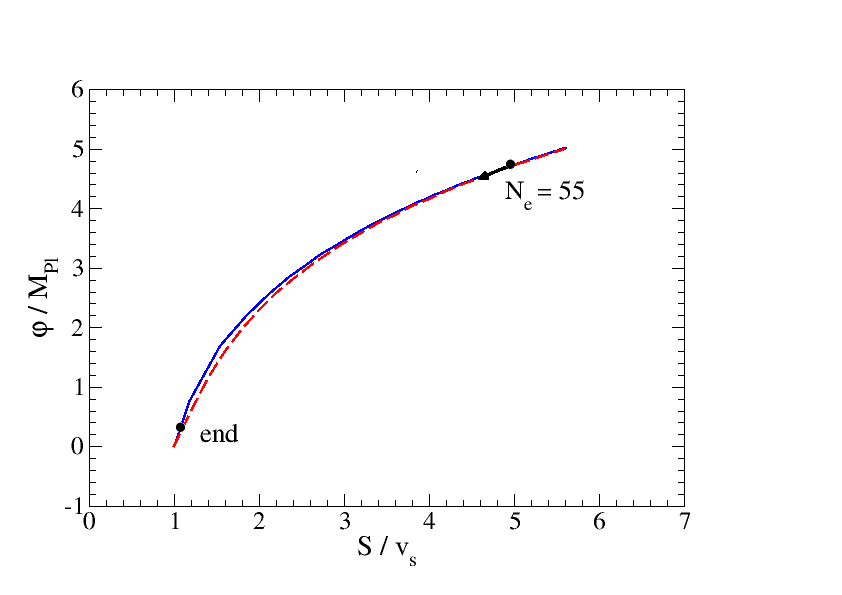}

\caption{Exact (blue) and valley (red) trajectories
for $S/v_S \le 5.6$, where 
$S_\text{end}/v_S =1.13$
and $S_*/v_S =5.05$ ($N=55.0$).
 }
\label{valley}
\end{center}
\end{figure}

By an exact solution we mean the solution obtained by 
the numerical integration of Eqs.~(\ref{eq-motion-phi}) and (\ref{eq-motion-S}).
In Figs.~\ref{s-phi} and \ref{valley} we show a representative 
 exact solution, where the input parameter values are:
 \begin{equation}
 y=0.0046\,, ~~\lambda_S=0.0114\,,~~\gamma=4.382\times 10^8\,,~~
 \beta =4680\,~~ (\bar{\gamma}=\gamma/\beta^2=20)\,,~~
 \label{bench4}
 \end{equation}
which give $\Lambda_H=5.89\times 10^{-2}M_\text{Pl}$ and
\al{
v_\sigma/M_\text{Pl} &=0.987\times 10^{-2}\,,\quad
v_S/M_\text{Pl}=1.46\times 10^{-2}\,,\quad U_0/M_\text{Pl}^4=-1.11\times 10^{-9}\,.
 \label{bench5}
}
The initial filed values at $t=t_0$ and  at the end of inflation 
for the trajectory (blue) plotted in Figs.~\ref{s-phi} and 
\ref{valley} are,
respectively,
 \begin{eqnarray}
 S_0/v_S &= &6.06\, ,~~~\varphi_0/M_\text{Pl}=5.38\,,
  \label{in-Value}\\
  S_\text{end}/v_S& =& 1.13\, ,~~~\varphi_\text{end}/M_\text{Pl}=0.601\,,
 \label{end-Value}
  \end{eqnarray}
and $ \dot{S}_0 =\dot{\varphi}_0=0$.
The left panel of Fig.~\ref{s-phi} shows the trajectory
at the beginning (i.e., $t\simeq  t_0$), from which we see that it is strongly oscillating, especially in the $S$ direction.
In the right panel we compare two trajectories with different initial values;
the blue one is for (\ref{in-Value}) and the red one is for 
 $S_0/v_S=6.06\, ,\varphi_0/M_\text{Pl}=5.36$.
We see  that the trajectories  converge very fast to a 
``fixed-point''  trajectory.
This fixed-point  solution (trajectory) is in fact very close to the approximate 
valley solution (\ref{phiv}). Fig.~\ref{valley}
 presents the exact solution (blue) and
the valley solution (red) in later time,
i.e., from just before the horizon exit to the end of inflation.
This is a desired feature on one hand, because, if 
it converges to a fixed-point solution
before the horizon exit, 
 the dependence of the initial field values at $t=t_0$
do not influence  the observable inflationary parameters.
\footnote{We have also considered the three-field system
(i.e., with $\sigma$ included) and 
 observed the same converging feature
of  trajectories in three dimensions.
}
On the other hand, this means that we have effectively 
a single-field system, so that we expect that the non-Gassianity
of the primordial  curvature perturbations will be 
very small \cite{Maldacena:2002vr,Creminelli:2004yq}.
\footnote{See Ref. \cite{Choudhury:2023kdb} for an exception.}
This is a prediction of our model.
In the subsequent sections  we will verify explicitly this expectation
without using the full valley approximation (a single-field approximation).

\begin{figure}[ht]
\begin{center}
\includegraphics[width=7cm]{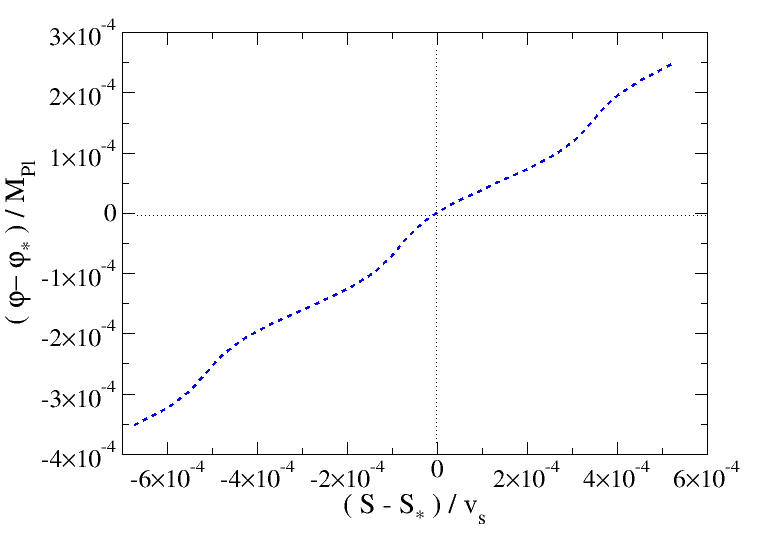}
\includegraphics[width=7cm]{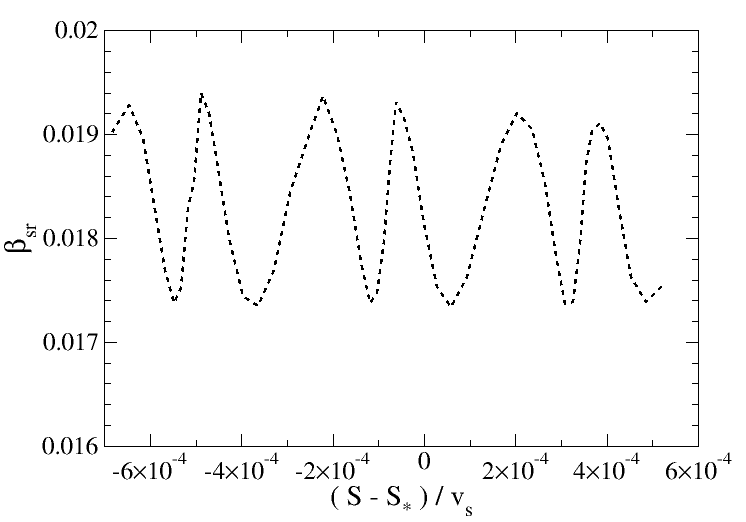}
\caption{Left: Zoomed trajectory in the $S$-$\varphi$ plane
near  $S=S_*$ (i.e., $t=t_*$). It is still slightly oscillating.
Right: $\beta_\text{sr}$ against 
$(S-S_*)/v_S$ near  $t=t_*$. This shows that 
$\beta_\text{sr}\ll 1$ is satisfied.
}
\label{oscillation}
\end{center}
\end{figure}

\begin{figure}[ht]
\begin{center}
\includegraphics[width=7cm]{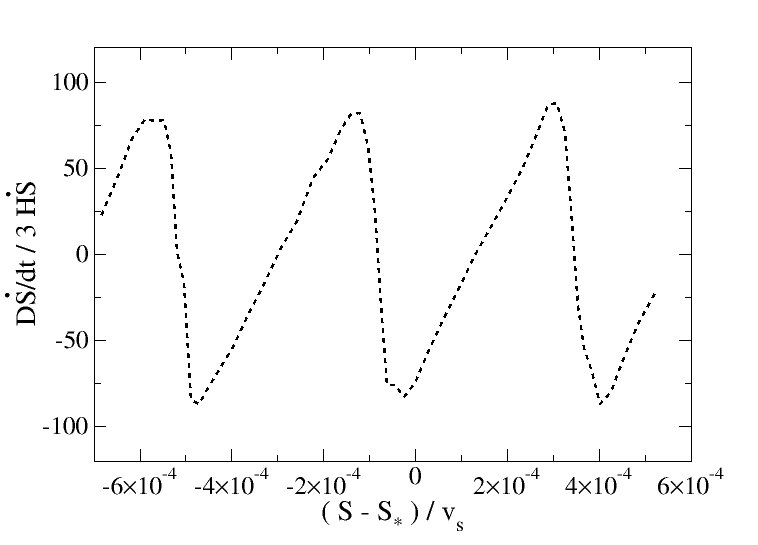}
\includegraphics[width=7.4cm]{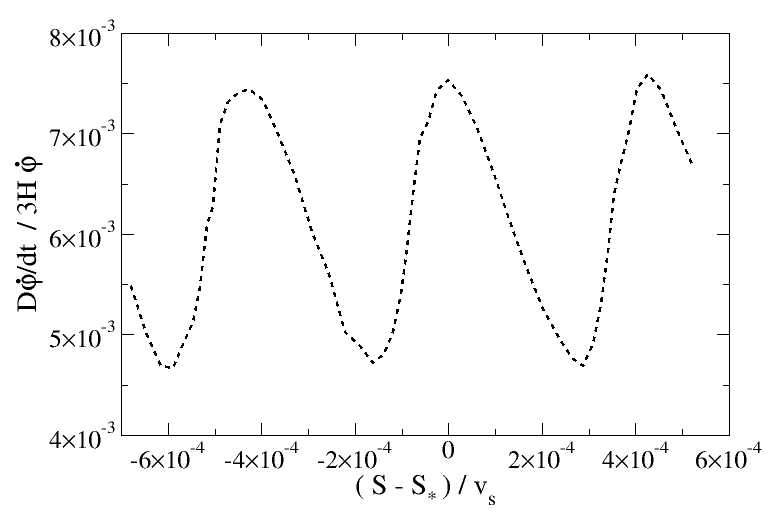}
\caption{Slow-roll condition (\ref{slow-roll1})
near $t=t_*$.
$({\cal D}\dot{S}/dt)/ (3 H \dot{S})$ (left)
and  $({\cal D}\dot{\varphi}/dt)/ (3 H \dot{\varphi})$ (right) are plotted
against $(S-S_*)/v_S$. In this example, $S$ is basically $\phi^\perp$,
where the $\phi^\perp$  direction 
 is perpendicular to the river of the valley.
}
\label{slow-roll}
\end{center}
\end{figure}

However,
there is a complication 
in computing the inflationary parameters using 
the $\delta N$ formalism \cite{Sasaki:1995aw,Nakamura:1996da,Sasaki:1998ug,Lyth:2004gb,Lyth:2005fi,Sugiyama:2012tj}, which  we will  point out below,
and we will come back to this problem when computing
the non-Gaussianity of the primordial  curvature perturbations explicitly in 
\ref{observables}.
The valley structure can be quantitated by
\begin{equation}
\eta_{IJ}
=M_\text{Pl}^2\,\frac{{\cal D}_I  V_{,J}}{V}\,,
\label{etaIJ}
\end{equation}
which is 
 a mass matrix (Hessian) normalized by $V/M_\text{Pl}^2$.
Therefore, for
a  prototype of the river-valley like potential,
there is a small as well as a large eigenvalue   of $\eta_{IJ}$;
the small one is the second order change of the potential 
 along the river at the bottom of the valley,  and the large one is that 
of the direction normal  to the river.
\footnote{
There is an interesting application of this fact
on a two-stage inflation \cite{Sasaki:1986hm}.}
 Note that $\eta_{IJ}$ is  also a slow-roll parameter. 
 This means that the slow-roll condition on the river-valley like  potential 
 is necessarily violated 
in  the direction normal  to the river.
As a result,
 the second time derivative of  $\phi$ in that direction,
  $\ddot{\phi}^{\perp}$, can be  
 large. That is,
  \begin{equation}
 \frac{ {\cal D}\dot{\phi}^{\perp}}{dt} 
 =\ddot{\phi}^{\perp}+
 \Gamma_{JK}^{\perp}\,\dot{\phi}^J
 \dot{\phi}^K
 \ll 3 H   \dot{\phi}^{\perp}
 \label{slow-roll1}
   \end{equation}
 is violated. Equivalently,
 the approximate equation
 \begin{equation}
 \dot{\phi}^{\perp}\simeq \frac{\,{\cal G}^{^{\perp}J}V_{,J}}{3H}
  \label{slow-roll2}
 \end{equation}
 can not be used to express   $\dot{\phi}^{\perp}$  in terms of $\phi^I$.
 The fact that   $\ddot{\phi}^{\perp}$ as well as a mass eigenvalue is large
 means that the background trajectory is oscillating 
 while slowly rolling down along the river.
 Note that the violation of (\ref{slow-roll1}) and (\ref{slow-roll2}) does not contradict 
 with $\epsilon\, ,\beta_\text{sr} \ll 1$, 
 where $\epsilon$ and $\beta_\text{sr}$
 are give in Eqs.~(\ref{epsilon3}) and (\ref{beta-sr}).
 The main reason for this is that  $\dot{\phi}^{\perp}$
 can be much smaller than RHS of Eq.~(\ref{slow-roll2}), such that
 $\dot{\phi}^{\perp}\ddot{\phi}^{\perp}$ in $\beta_\text{sr}$ can be small
 even if $\ddot{\phi}^{\perp}$ is large  \cite{Karciauskas:2018urs}.
 In the following figures, using the  representative example (\ref{bench4})
 of the set of parameters,
  we demonstrate  the situation described above.
  If we assume that $N=55$ for the representative 
  example (\ref{bench4}), then the filed values and their derivatives at 
  $t=t_*$
  should be
  \begin{eqnarray}
  S_*/v_S& =&   S(t=t_*)/v_S=5.05\, ,
  \varphi_*/M_\text{Pl}=  \varphi(t=t_*)/M_\text{Pl}=4.76\,,
    \label{star-Value}\\
   \dot{S}_*/v_S M_\text{Pl}& =&  
     -1.81\times 10^{-7}\, ,\dot{\varphi}_*/M^2_\text{Pl}
    =-5.30\times 10^{-6}\,.
  \nonumber
  \end{eqnarray}
 In the left panel of Fig.~\ref{oscillation} we show the zoomed trajectory 
 near $S=S_*$ (i.e., $t=t_*$). We see that the trajectory 
 is still slightly oscillating. In the right panel, the slow roll parameter
 $\beta_\text{sr}$ (\ref{beta-sr}) is plotted,
 which clearly shows that $\beta_\text{sr}\ll 1$ is satisfied.
 We have checked that this  is satisfied not only
 for this period, but also for all the time during inflation.
 The second derivative of $\phi^I$ with respect time
 in the period 
near $t=t_*$ is plotted in Fig.~\ref{slow-roll}, where the left panel is for 
 $S$ and the right one is for $\varphi$.
 We see from the left panel that the condition (\ref{slow-roll1}) is clearly 
 violated in the  $S$ direction, while it is satisfied in the $\varphi$
 direction.

A consequence of the river-valley like potential is
that we can not eliminate $\dot{\phi}^{\perp}$
using Eq. (\ref{slow-roll2}), which has an implication in calculating
the derivatives of $N$ with respect $\phi^{\perp}$
in the $\delta N$ formalism. How we overcome this problem 
will be explained in the subsections \ref{observables} and 
\ref{derivatives}.

Before we close this subsection we make a following remark about 
the applicability
of the NJL theory as an effective theory. 
This subject has been briefly touched 
in \ref{NJL}. Using concrete numbers given 
in (\ref{bench5})- (\ref{end-Value}) and also  (\ref{star-Value}) 
for the benchmark point (\ref{bench4}), we find
that $v_\sigma/\Lambda_H < 1,~
v_S/\Lambda_H < 1,~
(y S_0,~ y S_*)/\Lambda_H < 0.1$ are all satisfied.
We have made this consistency check for other sets of parameters
and found that the above requirement is satisfied for all the cases.

\subsubsection{$\delta N$ formalism}
Here we briefly summarize  the basic ingredients of the $\delta N$ formalism
\cite{Sasaki:1995aw,Nakamura:1996da,Sasaki:1998ug,Lyth:2004gb,Lyth:2005fi}, and we follow Ref. \cite{Sugiyama:2012tj} below.
We assume that the unperturbed Universe can be described by the
Friedmann-Lema\^itre-Robertson-Walker metric
\begin{equation}
ds^2 = -dt^2 +a(t)^2 d{\bf x}^2\,,
\end{equation}
which will be  perturbed.
The metric of the perturbed Universe in  the 
Arnowitt-Deser-Misner (ADM) form is given by
\begin{equation}
ds^2 = -\alpha^2 dt^2 +\gamma_{ij} (dx^i +\beta^i dt)(dx^j +\beta^j dt)\,,
\end{equation}
where $\gamma_{ij}$ can be written as
\begin{equation}
\gamma_{ij} =a^2 e^{2 \psi} \big( e^{h^{(T)}}\big)_{ij}\,,
\label{gammaij}
\end{equation}
and $\psi$ is  the spatial curvature perturbation.
We assume that  the scalar and vector perturbations in the
tensor perturbations are eliminated (gauged away) so that
$h^{(T)}_{ij}$ is traceless and transverse.
A physical quantity, relevant to our purpose, is the
curvature perturbation on constant density slice, $\zeta$.
 In the language of Ref. \cite{Sugiyama:2012tj}, $\zeta$ is $\psi$ in the 
 uniform-density gauge, that is, $\left.\psi\right|_\text{UD}=\zeta$.
 The perturbed system can be analyzed by expanding the non-linear field equations
 in the number of spatial derivatives, which corresponds to the expansion
 in $\xi\,(\ll 1)$, where $\xi$ is
the ratio of the comoving wavenumber $k$ and the comoving Hubble scale
$a H $.
It has been  shown in Ref. \cite{Sugiyama:2012tj}
 that,  up to and including  the next-to-leading
 order in the gradient expansions in the flat gauge,
 the perturbed scalar fields $\phi^I$ satisfy  the same equations of motion
 as the   unperturbed ones $\bar{\phi}^I$, 
 where $\psi$ in the flat gauge vanishes, $\left.\psi\right|_\text{F}=0$.
 This is regarded as a justification of the assumption of the separate Universe approach
(see Ref. \cite{Wands:2000dp} and references cited therein) that each area in the perturbed Universe, flatted  in a superhorizon scale,
 independently develops as an unperturbed Universe.
 
 Since $\left.\psi\right|_\text{F}=0$, one performs
 a gauge transformation from the flat gauge to the uniform-density gauge
to obtain the curvature perturbation on constant energy density slice,
 i.e., $\left.\psi\right|_\text{F}=0 \to 
 \left.\psi\right|_\text{UD}=\zeta$.
 The gauge transformation  is  a change of the time coordinate $t$,
 $t\to \hat{t}=t+\delta \hat{t}$,
 and it can be shown \cite{Sugiyama:2012tj} that the spatial metric 
 $\gamma_{ij}$ (\ref{gammaij}) transforms as a scalar up to and including
  the next-to-leading  order in the gradient expansions. 
  Therefore, 
  \begin{equation}
 \left. \gamma_{ij}\right|_\text{F}(t, {\bf x})
 =a^2(t)   \left.\big( e^{h^{(T)}}\big)\right|_\text{F}=  \left.\gamma_{ij}
 \right|_\text{UD}(\hat{t}, {\bf x})=
 a^2(\hat{t}) e^{2\left.\psi\right|_\text{UD}(\hat{t}, {\bf x})}
  \left.\big( e^{h^{(T)}}\big)\right|_\text{UD}\,,
 \end{equation}
 from which one finds
   \begin{equation}
   \zeta(\hat{t},{\bf x})=\left.\psi\right|_\text{UD}(\hat{t}, {\bf x})
   =\ln \Big( \frac{a(t)}{a(\hat{t})} \Big)
   =N-\hat{N}=\delta N(\hat{t},{\bf x})\,,\label{zeta}
     \end{equation}
 where
      \begin{equation}
     N =\int_{t_*}^{t}H (t') dt'\, ,  
         ~~~ \hat{N} =\int_{t_*}^{\hat{t}}H (t') dt'\, .
          \label{e-Fold4}
 \end{equation}
 
 Interestingly, 
$ \zeta=\delta N$ can be related to
 scalar fiels perturbations \cite{Sasaki:1995aw,Nakamura:1996da,Sasaki:1998ug,Lyth:2004gb,Lyth:2005fi},
 as we will see below.
 Once again, using the fact that 
 the perturbed scalar fields $\phi^I$  satisfy the same equations of motion as the unperturbed ones $\bar{\phi}^I$, we can express the energy density 
 in a perturbed system  from that in the unperturbed system.
 So, we perturb the homogeneous background solution in the unperturbed system
 by changing the initial field values $\bar{\phi}^I_*=\bar{\phi}^I(t_*)$ to
 a $\bf x$ dependent ones, i.e, $\bar{\phi}^I_*\to 
 \bar{\phi}^I_*({\bf x})=\bar{\phi}^I_*+
 \delta\phi^I_*({\bf x})$ at $t=t_*$. This scalar field perturbation makes
 the energy density becomes ${\bf x}$  dependent:
 $\rho(t, {\bf x})$ is  nothing but
  $\bar{\rho}(t,  \bar{\phi}^I_*({\bf x}))$ in the flat gauge, where 
  $\bar{\rho}(t, \bar{\phi}^I_*({\bf x}))$ is the energy density
  computed in the unperturbed system with the scalar perturbations
  $\delta\phi^I_*({\bf x})$ at $t=t_*$.
  In writing the perturbed energy density as $\bar{\rho}(t, \bar{\phi}^I_*({\bf x}))$, 
it is emphasized that the spatial inhomogeneity (${\bf x}$
dependence) comes solely from the initial values which depend on ${\bf x}$.
To be more explicit, we solve the equations of motions
  (\ref{EQM3})  with 
  $\bar{\phi}_*^I({\bf x})=\bar{\phi}_*^I+\delta\phi_*^I({\bf x})$
  for each ${\bf x}$.
  Then using $\rho=H^2/3 M_\text{Pl}$ and Eq.~(\ref{Hubble3}),
  we can compute the inhomogeneous $\bar{\rho}(t,  \bar{\phi}^I_*({\bf x}))$
  in the unperturbed system.

   Next, under the gauge transformation from the flat gauge to the uniform-density gauge,
 which is a change of the time coordinate $t\to \hat{t}=t+\delta \hat{t}$,
 the energy density behaves as a scalar, that is,
 \begin{equation}
 \bar{\rho}(t,  \bar{\phi}^I_*({\bf x}))
 = \hat{\rho}(\hat{t},  \bar{\phi}^I_*({\bf x}))=\bar{\rho}_0\,.\label{rhobar1}
  \end{equation}
  Since $\bar{\rho}_0$ on RHS is a constant independent of $t$, the $t$
  in LHS can not be arbitrary chosen.
 That is,  using the e-foldings $N$ instead of $t$, Eq.~(\ref{rhobar1}) 
   can be written as
 $\bar{\rho}(N,  \bar{\phi}^I_*({\bf x}))
= \bar{\rho}_0$, so  that for  a given initial $
\bar{\phi}^I_*({\bf x})$ with a fixed  $\bar{\rho}_0$
there exits a unique $N$, meaning that $N$ 
can be regarded as a function of $\bar{\phi}^I_*({\bf x})$
and consequently becomes ${\bf x}$ dependent.
\footnote{Here $\bar{\rho}_0$ plays the  role of  a time coordinate.}
Because of Eq.~(\ref{Hubble3}) the functional form of $N$
for each ${\bf x}$  is the same as $\hat{N}$
so that we may write
$N=\hat{N}(\bar{\phi}^I_*({\bf x});  \bar{\rho}_0)$ with
$\hat{N}(\bar{\phi}^I_*,\bar{\rho}_0 )=\hat{N}$,
arriving at
\cite{Sasaki:1995aw,Nakamura:1996da,Sasaki:1998ug,Lyth:2004gb,Lyth:2005fi}
\begin{equation}
 \zeta({\bf x})=\delta N({\bf x})
 =\hat{N}(\bar{\phi}^I_*+\delta \phi^I_*({\bf x});\bar{\rho}_0) -\hat{N}=
N_I\, \delta \phi^I_*({\bf x})+
 \frac{1}{2!}N_{IJ}\, \delta \phi^I_*({\bf x})\delta \phi^J_*({\bf x})\cdots\,,
 \label{zeta2}
  \end{equation}
  where $\cdots$ stands for higher order derivatives, and
   \begin{equation}
   N_I = \frac{\partial \hat{N}(\bar{\phi}^K_*,\bar{\rho}_0 )}{\partial \bar{\phi}_*^I}\,,
      ~~~N_{IJ}=\frac{\partial^2 \hat{N}(\bar{\phi}^K_*,\bar{\rho}_0 )}{\partial \bar{\phi}^I_* \partial \bar{\phi}_*^J} \,,\cdots
    \end{equation}
  
  Before we come to discuss inflational observables, we follow Refs. \cite{Gong:2011uw,Elliston:2012ab,Kaiser:2012ak}
  to express  $ \delta \phi^I_*({\bf x})$ in terms of a covariant quantity ${\cal Q}^I$,
reflecting the fact that
the background system is described by the covariant equations of motion
 (\ref{EQM3}). To this end, we note that there exists a unique geodesic,
  parametrized by $\lambda$, 
 in the field space that links $\bar{\phi}^I_*$ and  $\bar{\phi}^I_*({\bf x}) $.
The geodesic is described by  $\bar{\phi}^I_*(\lambda)$
which satisfies ${\cal D}_\lambda \,(d \bar{\phi}^I_*(\lambda)/d \lambda)=0$ with
 $\bar{\phi}^I_*(\lambda=0) =\bar{\phi}^I_*$ and  $\bar{\phi}^I_*(\lambda=\lambda_0)=
  \bar{\phi}^I_*({\bf x})$. Then we have
   \begin{eqnarray} 
   \delta \phi^I_*({\bf x})&=&
     \bar{\phi}^I_*({\bf x})- \bar{\phi}^I_*=
     \left.\frac{d \bar{\phi}^I_*(\lambda)}{d \lambda}\right|_{\lambda=0}\epsilon
     + \frac{1}{2!} \left.\frac{d^2 \bar{\phi}^I_*(\lambda)}{d \lambda^2}
     \right|_{\lambda=0}\epsilon^2+O(\epsilon^3)\nn\\
     &=&{\cal Q}^I \epsilon -\frac{1}{2!}\Gamma_{JK}^I{\cal Q}^J{\cal Q}^K\epsilon^2
     +O(\epsilon^3)\nn\\
      &=&{\cal Q}^I  -\frac{1}{2!}\Gamma_{JK}^I{\cal Q}^J{\cal Q}^K
     +O({\cal Q}^3)\,,\label{Qyu}
 \end{eqnarray}    
  where we have identified
 ${\cal Q}^I$ with 
  the tangent vector $d \bar{\phi}^I_*(\lambda) /d \lambda$ at $\lambda=0$,
 the geodesic equation $ 
{\cal D}_\lambda \,(d \bar{\phi}^I_*/d \lambda)=0$ is used to arrive at the second equation,
and  $\epsilon$ is absorbed into ${\cal Q}^I$ to obtain the last line.
Using  ${\cal Q}^I$ the curvature perturbation (\ref{zeta2}) can now be written as
\cite{Elliston:2012ab,Kaiser:2012ak}
 \begin{equation}
 \zeta({\bf x}) =N_{I}\,{\cal Q}^I({\bf x})+\frac{1}{2}
( {\cal D}_I {\cal D}_J N){\cal Q}^I({\bf x}){\cal Q}^J({\bf x})+   O({\cal Q}^3)\,.
\label{zeta3}
 \end{equation}

\subsubsection{Inflationary observables}\label{observables}
As we have seen in the previous subsection,
the spatial inhomogeneity (${\bf x}$ dependence) of $\zeta$ in the $\delta N$ formalism
originates from the inhomogeneity of the initial field values at  horizon exit,
i.e., $t=t_*$.  
Inflationary observables 
can be calculated from 
correlation functions of $\zeta$ \cite{Bartolo:2004if}.
Our interest here is focused on 
the power spectrum and bispectrum
\cite{Komatsu:2001rj,Maldacena:2002vr} 
in the wave number space, $\langle \,\zeta({\bf k}_1) \,\zeta ({\bf k}_2)\,\rangle$ 
 and 
$\langle \,\zeta({\bf k}_1) \,\zeta ({\bf k}_2)\,\zeta ({\bf k}_3)\,\rangle$,
respectively, where
$\zeta({\bf k})=\int  d^3{\bf k}/(2\pi)^3 e^{ -i {\bf k}\cdot {\bf x}}\zeta({\bf x})$
is the Fourier component of $\zeta({\bf x})$, and 
the expectation values are calculated with respect to  
the Bunch-Davies Vacuum.
As we see from Eq. (\ref{Qyu}), 
RHS of Eq. (\ref{zeta3}) is an expansion in ${\cal Q}$.
Accordingly, we restrict ourselves to  the power spectrum and bispectrum
in the lowest nontrivial order  
in ${\cal Q}$ \cite{Elliston:2012ab,Kaiser:2012ak}:
\begin{eqnarray}
\langle \,\zeta({\bf k}_1) \,\zeta ({\bf k}_2)\,\rangle
&=&(2\pi)^3 \delta^3({\bf k}_1+{\bf k}_2)\frac{2\pi^2}{k_1^3} {\cal P}_\zeta (k_1)
\nn\\
&=&N_I N_J \langle \,{\cal Q}^I ({\bf k}_1) {\cal Q}^J({\bf k}_2) \rangle\,,
\\
\langle \,\zeta({\bf k}_1) \,\zeta ({\bf k}_2)\,\zeta ({\bf k}_3)\,\rangle
&=&(2\pi)^3 \delta^3({\bf k}_1+{\bf k}_2+{\bf k}_3)B_\zeta (k_1,k_2,k_3)\nn\\
& = &N_I N_J N_K \langle \,{\cal Q}^I ({\bf k}_1) {\cal Q}^J({\bf k}_2){\cal Q}^K
 ({\bf k}_3)\rangle\nn\\
 &+&N_I N_J {\cal D}_K{\cal D}_L N
 \int\frac{d^3 {\bf q}}{(2\pi)^3} 
 \langle {\cal Q}^K({\bf k}_1-{\bf q}) {\cal Q}^I ({\bf k}_2)\rangle
 \,  \langle {\cal Q}^L({\bf q}) {\cal Q}^J({\bf k}_3)\rangle\,
 \label{three-point}
\end{eqnarray}
with $k=|{\bf k}|$.

To proceed, we recall that the non-Gaussianity parameter $f_{NL}$ is defined as \cite{Komatsu:2001rj,Maldacena:2002vr}
\begin{equation}
f_{NL} =\frac{5}{6}\frac{B_\zeta (k_1,k_2,k_3)}{P_\zeta{k_1}+
P_\zeta{k_2}+\mbox{c.p.}}\,,
\end{equation}
where c.p. stands for cyclic permutation of $k_1,k_2$ and $k_3$.
Further, the Mukhanov-Sasaki equation
\cite{Mukhanov:1988jd,Sasaki:1995aw} in the case of   multi-component systems
contains a matrix-valued parameter \cite{Nakamura:1996da}
(see also Ref. \cite{Elliston:2012ab})
\begin{equation}
\epsilon_{IJ}=
-G_{IJ}\frac{\dot{H}}{H^2}+\Big(\frac{G_{IK}G_{JL}}{M^2_\text{Pl}}-\frac{1}{3}R_{IKJL}\Big)
\frac{\dot{\phi}^K\dot{\phi}^L}{H^2}-\frac{{\cal D}_I{\cal D}_J V}{3H^2}
\label{epsilonIJ}\,,
\end{equation}
which is slowly varying in time during inflation.
We assume that $\epsilon_{IJ*}=\epsilon_{IJ}$ at the horizon exit $t=t_*$ remain 
constant during inflation
 (as it is  assumed   in  Ref. \cite{Nakamura:1996da}) and 
 also that the first term on RHS of
 Eq. (\ref{three-point}), 
 the three-point correlation function,
 is very small \cite{Seery:2005gb,Lyth:2005qj}.
 Under these assumptions, we
 obtain \cite{Sasaki:1995aw,Nakamura:1996da,Lyth:2005fi,Seery:2005gb,Lyth:2005qj,Elliston:2012ab,Kaiser:2012ak,Bartolo:2004if}
\begin{eqnarray}
n_s &=&\left.1+\frac{d \ln {\cal P}_\zeta(k)}{d \ln k}\right|_{k=a_* H_*}
=1-2\frac{\epsilon_{IJ*}N^I N^J}{N_K N^K } \,,\label{ns30}\\
r &=&\frac{8}{M_\text{Pl}^2 N_I N^I}\,,~~~A_s =(H_*/(2\pi)^2)\,N_I N^I\,,
\label{rs30}\\
f_{NL} &=&\frac{5}{6}\frac{({\cal D}_I {\cal D}_J N) N^I N^J}{(N_K N^K)^2}\,,
\label{fNL30}
\end{eqnarray}
where
 $1/a_* H_*$ is the comoving Hubble radius
 at the horizon exit.

We may now proceed with the application of the general formulae
Eqs.~(\ref{ns30})-(\ref{fNL30}) to our concrete system, 
where the evolution of the background Universe is described by
a heavy as well as  a light mode. To obtain reliable results, we therefore have to incorporate
the fact that
the fluctuations of the heavy scalar mode do not produce perturbations on cosmologically relevant scales \cite{Pilo:2004ke} 
(see also Ref.~\cite{Bartolo:2004if}).
To this end, we first define the light and heavy modes,
$\bar{\phi}_L$ and $\bar{\phi}_H$,  by diagonalizing
$\epsilon_{IJ*}$:
\begin{equation}
R\, \epsilon_{*} R^T =\mbox{diag.}(\, \epsilon_H\,, \,\epsilon_L\,)\,,\,
\mbox{where}~\epsilon_L \ll \epsilon_H~\mbox{and}~R=
\left(\begin{array}{cc}
\cos \theta &- \sin \theta\\
\sin \theta & \cos \theta
\end{array}\right)\,,
\end{equation}
which means that the  original background fields $\bar{\varphi}$
and $\bar{S}$ are written as
$\bar{\varphi}=\cos\theta\, \bar{\phi}_H+\sin\theta \bar{\phi}_L\,,\,
\bar{S}=-\sin\theta \bar{\phi}_H+\cos\theta \bar{\phi}_L$.
 Then we perturb the background  only in  the light mode direction.
 That is,
we consider only the change of the initial values of the form
 $\delta\phi_{L*}({\bf x})\neq 0$ but $\delta\phi_{H*}({\bf x})=0$, which
 can be achieved in our two-component system by imposing
$\delta \varphi_*({\bf x})/\delta S_*({\bf x})= \tan\theta$.
In this way,   the fluctuations from the 
heavy mode are suppressed at the horizon exit,
and we may assume that they remain suppressed,
because $\epsilon_{IJ}$ is only slowly changing during inflation.
The discussions above then lead us to
\begin{equation}
\zeta({\bf x})
=\delta N({\bf x})=N'\,\delta\phi_{L*}({\bf x})+\frac{1}{2!}N''\,
(\delta\phi_{L*}({\bf x}))^2+\cdots\,,
\end{equation}
where
\begin{eqnarray}
N' &=& \frac{d\hat{N}(\bar{\phi}_{L*} ,\bar{\rho}_0)}{d\bar{ \phi}_{L*}}=
N_\varphi\, \sin\theta +N_S\, \cos\theta \,, \nn\\
N'' &=&\frac{d^2\hat{N}(\bar{\phi}_{L*} ,\bar{\rho}_0)}{d\bar{ \phi}_{L*}^2}=
N_{\varphi\varphi}\, \sin^2 \theta +N_{SS}\, \cos^2\theta +
2N_{\varphi S}\, \sin\theta\cos\theta \,.
\end{eqnarray}

With 
$\phi_{L*}$ and $  \phi_{H*}$, 
the ``mass term'' is  now diagonal,  but not their kinetic term in general.
However, in the present case, where the heavy mode $  \delta \phi_{H}$ remains 
suppressed,  the kinetic term for the light mode can be simply read off from
 $ G_{IJ} \dot{\phi}^I \dot{\phi}^J$:
\begin{eqnarray}
 G_{IJ} \dot{\phi}^I \dot{\phi}^J
  \to & &
   \big(G_{\varphi\varphi} \,\sin^2\theta+
    G_{SS} \,\cos^2\theta\big)\dot{\delta \phi}_L \dot{\delta\phi}_L\nn\\
 & =& \Big(\sin^2\theta+
   e^{-\sqrt{2/3} \varphi/M_\text{Pl} } \,\cos^2\theta\Big)
    \dot{\delta\phi}_L \dot{\delta\phi}_L\nn\\
 &=&   
     G_{L} \,\dot{\delta \phi}_L \dot{\delta \phi}_L\,,
\end{eqnarray}  
which, togehther with Eqs.~(\ref{ns30})-(\ref{fNL30}), gives
\begin{eqnarray}
n_s &=&
=1-\frac{2\epsilon_{L*}}{G_{L*}} \,,\label{ns3}\\
r &=&\frac{8G_{L*}}{M_\text{Pl}^2 (N')^2}\,,~~~
A_s =\frac{H_*}{(2\pi)^2}\,\frac{(N')^2}{G_{L*}}\,,
\label{rs3}\\
f_{NL} &=&\frac{5}{6}\frac{{\cal D}  N' }{(N')^2}\
=\frac{5}{6}\frac{ N''+N'
e^{-\sqrt{2/3} \bar{\varphi}_*}\cos^2\theta
 \sin\theta\,\big[\sqrt{6}M_\text{Pl}G_{L*}\big]^{-1} }{(N')^2}\,,
\label{fNL3}
\end{eqnarray}
where we have used:
$G'_L=d G_L/d \phi_L=-\sqrt{2/3} \cos\theta^2(\varphi' / M_\text{Pl})
e^{-\sqrt{2/3}\varphi /M_\text{Pl}}$
and $\varphi' =\sin\theta$.

\subsubsection{Numerical calculation of  the derivatives of $N$}
\label{derivatives}

To use $\delta N$ formalism, we have to compute
the derivatives of $\hat{N}(\bar{\phi}^I_*+\delta \phi^I_*({\bf x});
\bar{\rho}_0)$ with respect
to $\bar{\phi}_*^I$ at $  \delta \phi^I_*({\bf x})=0$.
This means that the derivatives $N_{I}\, ,N_{IJ}\, \dots$
are independent of ${\bf x}$.
Therefore, in the following discussion we simply denote  the function
by $N(\bar{\phi}^I_*)$, where we suppress also  ${\hat{}}$ on $N$ below.
In principle $N$ can depend also  on $\dot{\bar{\phi}}^I_*$,
which, however,  is usually assumed to be very small, and
the  $\dot{\bar{\phi}}^I_*$ dependence is indirectly included
by eliminating  $\dot{\bar{\phi}}^I$ trough the slow-roll equation
(\ref{slow-roll2}). In our case, where the slow-roll condition
 (\ref{slow-roll1})
in a certain direction is badly violated, it is not clear 
that the $\dot{\bar{\phi}}^I_*$ dependence is small, and  even if it is small,
we can not use Eq. (\ref{slow-roll2})  to eliminate  $\dot{\bar{\phi}}^I_*$ in that direction.

\begin{figure}[ht]
\begin{center}
\hspace{0.7cm}
\includegraphics[width=8cm]{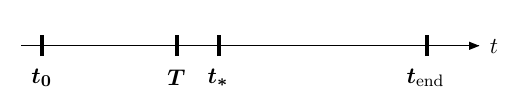}

\caption{The time axis. The background solution
$\bar{\phi}^I(t)$ 
 starts to run at $t_0$
with a vanishing velocity, i.e., $\dot{\bar{\phi}}^I(t_0)=0$.
A perturbed system is described by $\phi^I_T(t)$, which
is an exact solution with the initial values at $t=T$;
$\phi^I_T(T)=\bar{\phi}^I (T)+\Delta \phi_T^I$
and  $\dot{\phi}^I_T(T)=\dot{\bar{\phi}}^I(T)$. 
The variation from the background solution
 at the horizon exit ($t=t_*$)
can be numerically computed 
}
\label{time}
\end{center}
\end{figure}
First we describe how we eliminate  $\dot{\bar{\phi}}^I$.
The background solution is the one with the initial values set at 
$t=t_0$, which we denote by $\bar{\phi}^I(t)$.
In addition to  $\bar{\phi}^I(t)$, we introduce $\phi^I_T(t)$
which is also an exact solution with the initial value
$\phi^I_T(T)=\bar{\phi}^I (T)+\Delta \phi_T^I$
and $\dot{\phi}^I_T(T)=\dot{\bar{\phi}}^I(T)$, where
$t_0 \ll T < t_*$ as it is illustrated in Fig.~\ref{time}.
Then at $t=t_*$ we calculate $\Delta \phi_*^I$
and $\Delta\dot{ \phi}_*^I$
from
\begin{equation}
\Delta \phi_*^I
=\phi_T^I(t_*)-\bar{\phi}^I(t_*)\,,~~~
\Delta \dot{\phi}_*^I
=\dot{\phi}_T^I(t_*)-\dot{\bar{\phi}}^I(t_*)\,.
\label{Delta-phi}
\end{equation}
 Since $\phi_T^I(t)$ is an exact solution, 
there exist unique $\Delta \phi_*^I$ and 
$\Delta \dot{\phi}_*^I$ at $t=t_*$ for a given $\Delta \phi_T$.
 If we regard $\Delta \phi_*^I$ as independent,
$\Delta \dot{\phi}_*^I$ is no longer an independent variation.
Equivalently, we use
 $\dot{\bar{\phi}}_*^I=\dot{\phi}_T^I(t_*)$
to  substitute  Eq.~(\ref{slow-roll2}).
 
Note that we can not choose $\Delta \phi^I_*$ freely, 
because the initial  values of $\phi_T^I(t)$ are set at $t=T$. 
Therefore,  we introduce a set of
$\Delta_i \phi_T^I$ with $i=1,\dots, M$  and  then compute numerically
\begin{equation}
\Delta_i N=N(\bar{\phi}^I(t_*)+\Delta_i \phi_*^I)-N(\bar{\phi}^I(t_*))
\label{DN}
\end{equation}
for each $i$, where $\Delta_i \phi_*^I$ is calculated from Eq.~(\ref{Delta-phi}).
At the same time, we can expand RHS of Eq.~(\ref{DN}) in $\Delta_i \phi_T^I$:
\begin{equation}
\Delta_i N=N_{I}\Delta_i \phi_*^I+
\frac{1}{2!}N_{IJ}\Delta_i \phi_*^I\Delta_i \phi_*^J+
\frac{1}{3!}N_{ IJK}\Delta_i \phi_*^I\Delta_i \phi_*^J\Delta_i \phi_*^K+\cdots\,,
\label{linear-system}
\end{equation}
where 
$\Delta_i N$ and $\Delta_i \phi_*^I$
are explicitly  known, and
the unknowns are derivatives, $N_{I},N_{ IJ},\cdots$.
This means that, if we truncate the series at a certain  order,
 RHS of Eq.~(\ref{linear-system}) can define a  system of linear equations for
the unknowns.
For instance, to obtain the first and second derivatives, we need
five independent equations, 
because there are $3+2=5$ unknowns, $N_{S}\,,N_{\varphi}\,,
N_{SS}\,,
N_{S\varphi}$ and $N_{\varphi\varphi}$.
If we include the third derivatives, we need $5+4=9$ equations.

To use Eqs.~(\ref{ns3})-(\ref{fNL3}) we have to compute $N'$ and $N''$ numerically.
To this end,  we ``excite'' only the light mode $\delta\phi_L$,
which can be achieved by $\Delta \varphi_T=\sin\theta_T \Delta \phi_{LT}$ 
and $\Delta S_T=\cos\theta_T\Delta\phi_{LT}$,
where $\theta_T$ is calculated at $t=T$.
Then we obtain
\begin{eqnarray}
\Delta N &=&
N\big(\bar{\varphi}(t_*)+\Delta\varphi_*,
\bar{S}(t_*)+\Delta S_*\big)-N\big(\bar{\varphi}(t_*),\bar{S}(t_*)\big)\nn\\
&=&
N' \Delta \phi_{L*}+
\frac{1}{2!}N'' \big(\Delta \phi_{L*}\big)^2+
\frac{1}{3!}N''' \big(\Delta \phi_{L*}\big)^3+\cdots\,,
\label{linear-system-L}
\end{eqnarray}
where
$\Delta\varphi_*=\varphi_T(t_*)-\bar{\varphi}(t_*)=\sin\theta
\Delta\phi_{L*}\,,
\Delta S_*=S_T(t_*)-\bar{S}(t_*)=\cos\theta
\Delta\phi_{L*}$,
and we assume that $\theta$ does not change, i.e.,
$\theta_T=\theta$.

\begin{figure}[ht]
\begin{center}
\includegraphics[width=7cm]{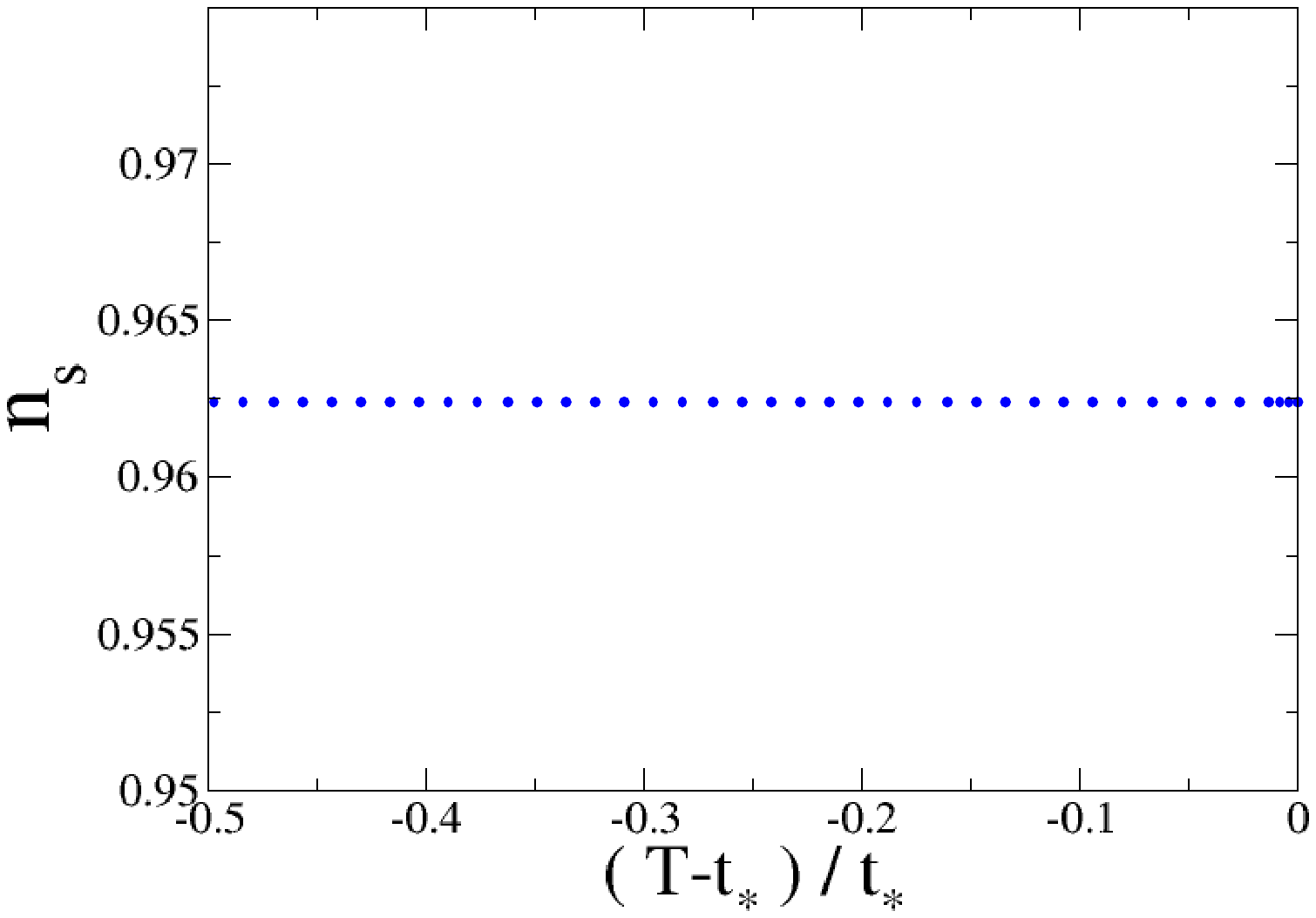}
\includegraphics[width=7cm]{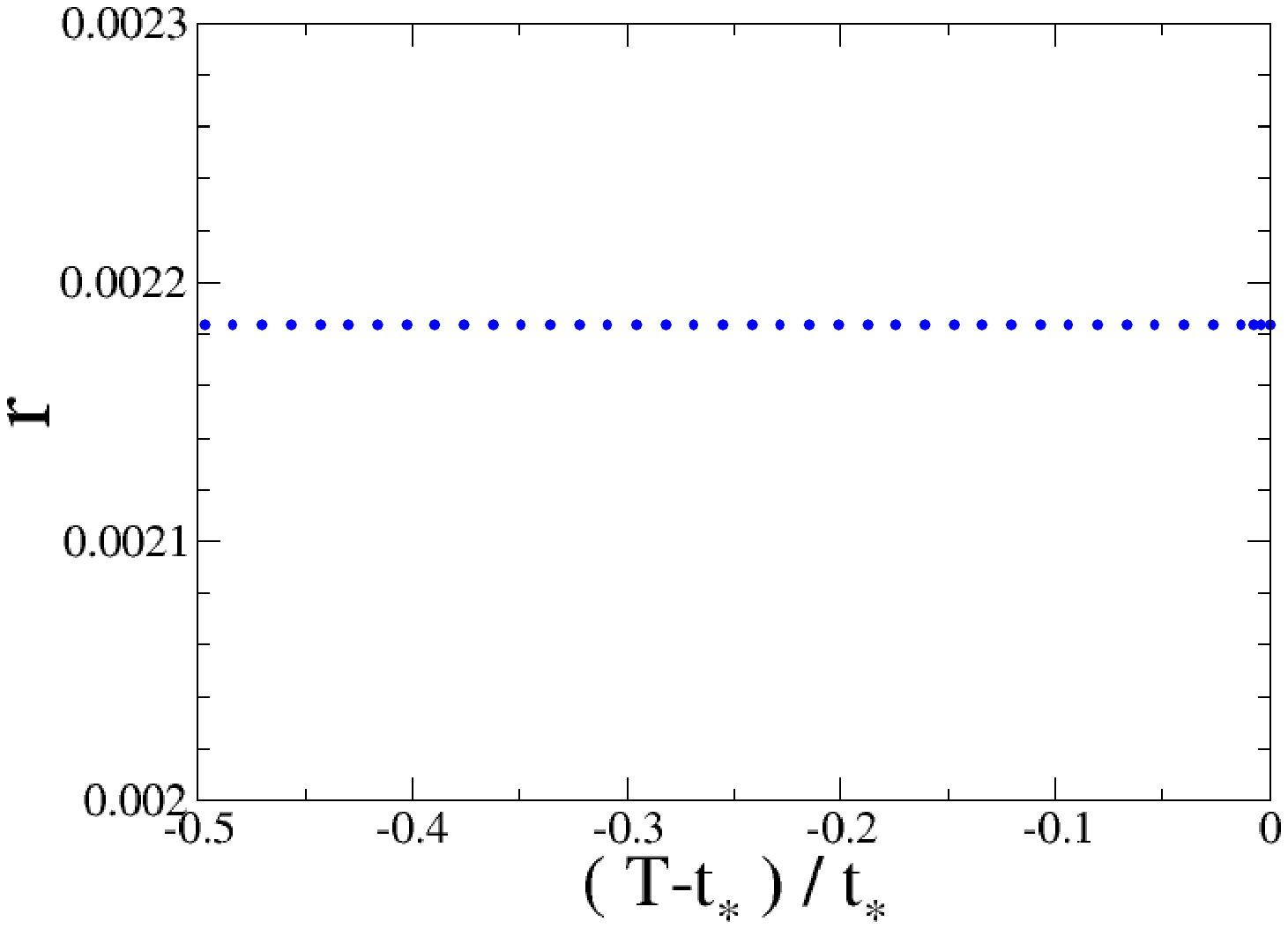}
\includegraphics[width=7cm]{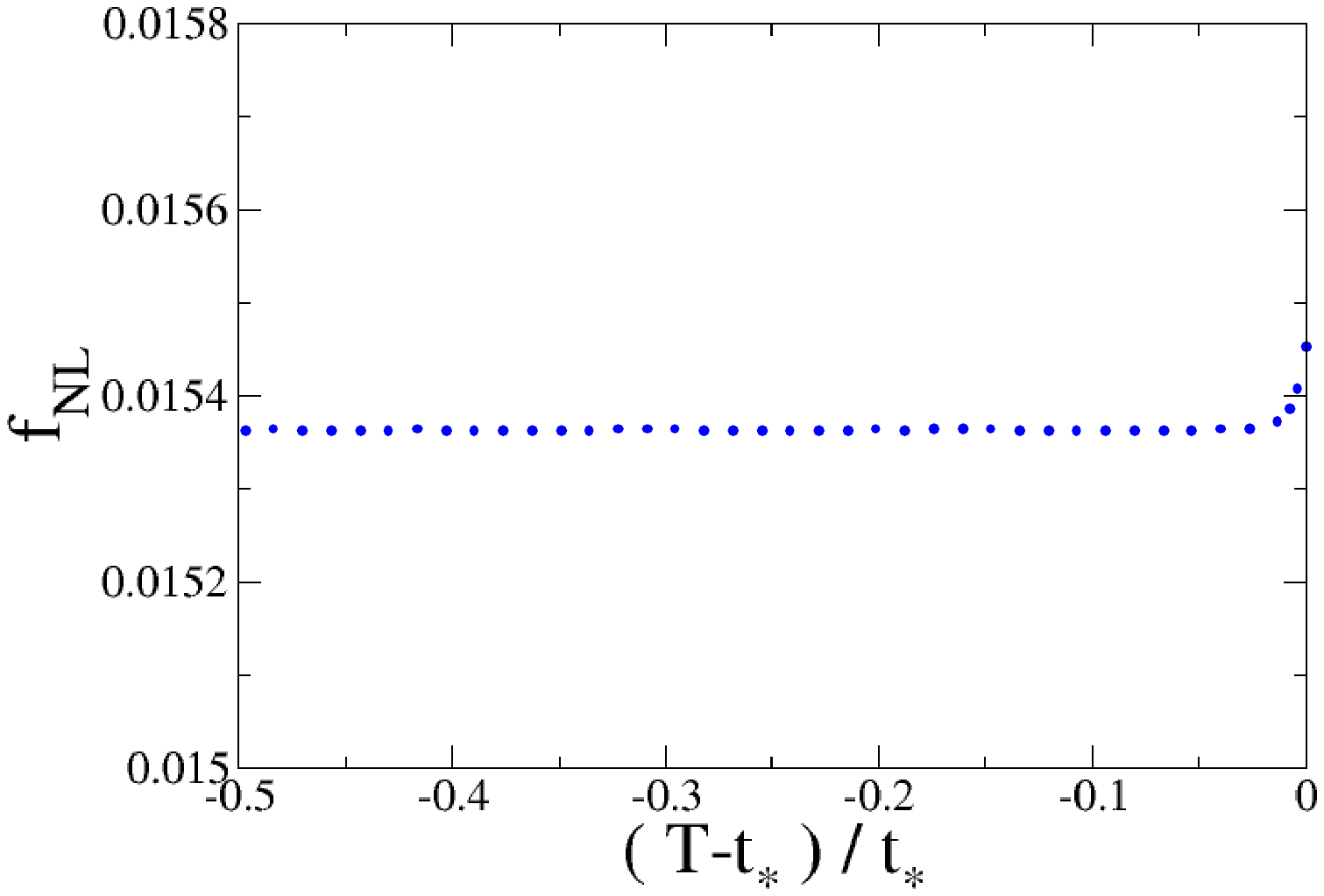}
\caption{
The spectral index $n_s$ (left), 
the tensor-to-scalar ratio $r$ (right)
and the non-Gaussianity parameter $f_{NL}$ (bottom)
 as a function of $(T-t_*)/t_*$.
}
\label{T-ns}
\end{center}
\end{figure}

Using the method described above we have computed the inflationary parameters,
which are shown in Fig.~\ref{T-ns}, where we have included 
fourth derivatives $N''''$ to improve the accuracy.
The spectral index $n_s$ (left), the tensor-to-scalar ratio $r$ (right)
and the non-Gaussianity parameter $f_{NL}$ (bottom), respectively,
are plotted as a function of $(T-t_*)/t_*$.
From these figures we may conclude that the $T$ dependence is very small.
Note that for  $T-t_*=0$ we have
$\Delta \dot{S}_*=\dot{S}_T(t_*)-\dot{\bar{S}}(t_*)=0$ and
$\Delta \dot{\varphi}_*=\dot{\varphi}_T(t_*)-
\dot{\bar{\varphi}}(t_*)=0$.
This means that  inflationary parameters
  depend  only slightly on
$\dot{\bar{S}}_*$ and  $\dot{\bar{\varphi}}_*$,
because 
$\Delta \dot{S}_*\neq 0$ and
$\Delta \dot{\varphi}_*\neq 0$ for $T < t_*$.
Therefore, in the following analysis we calculate the inflationary
parameters for $T=t_*$ to simplify the calculation.
Accordingly, for a given background trajectory $\bar{\phi}^I(t)$,
we perturb it by changing the initial values at $t=t_*$:
\begin{equation}
\bar{\phi}^I(t_*)=\bar{\phi}^I_*\to \bar{\phi}^I_*+\Delta \phi^I_*\, ,~~~
\dot{\bar{\phi}}^I(t_*)=
\dot{\bar{\phi}}^I_*\to \dot{\bar{\phi}}^I_*+\Delta \dot{\phi}^I_*
~~\mbox{with}~\Delta \dot{\phi}^I_*=0~(I=\varphi,S)\,,
\end{equation}
where
$\Delta \varphi_* =\sin\theta \Delta \phi_{L*}\,,
~\Delta S _* =\cos\theta \Delta \phi_{L*}$ and $ \Delta \dot{\phi}_{L*}=0$.

\subsubsection{Result}
\label{result}
We recall  that  the result depends on $y$ and $\lambda_S$  only slightly
in the valley approximation \cite{Aoki:2021skm} and 
may  assume that this will be also the case
in the present treatment.  In the following discussions we therefore
 fix them   at
 $y=0.0046\,, \lambda_S=0.0114$,
 as in (\ref{bench4}), which gives
 \begin{equation}v_S\simeq M_\text{Pl}/\sqrt{\beta} \,,
v_S/v_\sigma \simeq1.48\,,
 U_0/v_S^4\simeq-0.0244\,,
\end{equation}
and we are left with the free parameters $\beta$ and $\gamma$.
Furthermore, since the inflationay parameters, $n_s\,,r\,$ and
$\beta^2 A_s$, depend only on $\bar{\gamma}=\gamma/\beta^2$
in the valley approximation \cite{Aoki:2021skm},
we may assume that this is also the case here, although the potential 
$V(\phi^I)$  for our two-field system (\ref{eq:LEG}) depends explicitly on $\gamma$ and $\beta$.
The reason is that the trajectories fast converge  to a fixed-point trajectory,
as it is shown in Fig.~\ref{valley} (red), which is nearly the lowest part of
the  river-valley like potential (\ref{eq:LEG}) and that it depends only on 
$\bar{\gamma}$. As we  see from  TABLE \ref{table},
which shows  $n_s\,,r\,,f_{NL}$ and $\beta^2 A_s$
at $N=50\,,55$ and $60$ 
for $\beta= 4000\, ,4500$ and $5000$, respectively,
 with $\bar{\gamma}=20$ fixed,
this assumption is justified for these examples.
 Therefore, to simplify the calculation, we calculate the inflationary parameters 
first at fixed $\bar{\gamma}$ and $\beta$ and assume that 
the experimental constraint \cite{Aghanim:2018eyx,Planck:2018jri}
\al{
\ln (10^{10} A_s) = 3.044 \pm 0.014
\label{As_constraint}
}
can satisfied if we change $\beta$ appropriately without changing 
$\bar{\gamma}$.

\begin{table}
\begin{tabular}{|c|c|c|c|c|c|c|}
\hline
 &   $\beta$ &  $n_s$ & $r$ & $f_{NL}$ & $\beta^2 A_s$ & $\ln(10^{10} A_s)$\\ \hline
 $N=50$ & $\begin{array}{c}
4000\\4500\\5000
 \end{array}$ &$\begin{array}{c}
 0.958 \\ 0.958\\0.959
  \end{array}$& $\begin{array}{c}
 0.00273 \\ 0.00273\\ 0.00264
  \end{array} $& $\begin{array}{c}
 0.0173 \\ 0.0173\\0.0170
   \end{array} $& $\begin{array}{c}
0.0363\\0.0363\\0.0375
  \end{array} $& $\begin{array}{c}
3.12 \\2.89\\ 2.70
  \end{array}$\\ \hline
   $N=55$ & $\begin{array}{c}
4000\\4500\\5000
 \end{array}$ &$\begin{array}{c}
0.962 \\ 0.962\\0.962
  \end{array}$& $\begin{array}{c}
  0.00224 \\ 0.00224\\0.00218
  \end{array} $& $\begin{array}{c}
 0.0157 \\ 0.0157\\0.0155
  \end{array} $& $\begin{array}{c}
0.0443\\0.0443\\0.0456
  \end{array} $& $\begin{array}{c}
 3.32 \\3.09\\ 2.90
  \end{array}$\\ \hline
   $N=60$ & $\begin{array}{c}
4000\\4500\\5000
 \end{array}$ &$\begin{array}{c}
0.965\\ 0.965\\0.966
  \end{array}$& $\begin{array}{c}
0.00187 \\ 0.00187\\ 0.00182
  \end{array} $& $\begin{array}{c}
0.0144 \\ 0.0144\\0.0142
  \end{array} $& $\begin{array}{c}
0.0531\\0.0531\\0.0546
  \end{array} $& $\begin{array}{c}
 3.50 \\3.27\\ 3.08
  \end{array}$\\ \hline

\end{tabular}
\caption{$\beta$ (in)dependence of $n_s$, $r$, $f_{NL}$ and $\beta^2 A_s$ for 
a fixed  $\bar{\gamma}=\gamma/\beta^2$ at $=20$.}
\label{table}
\end{table}

\begin{figure}[ht]
\begin{center}
\includegraphics[width=14cm]{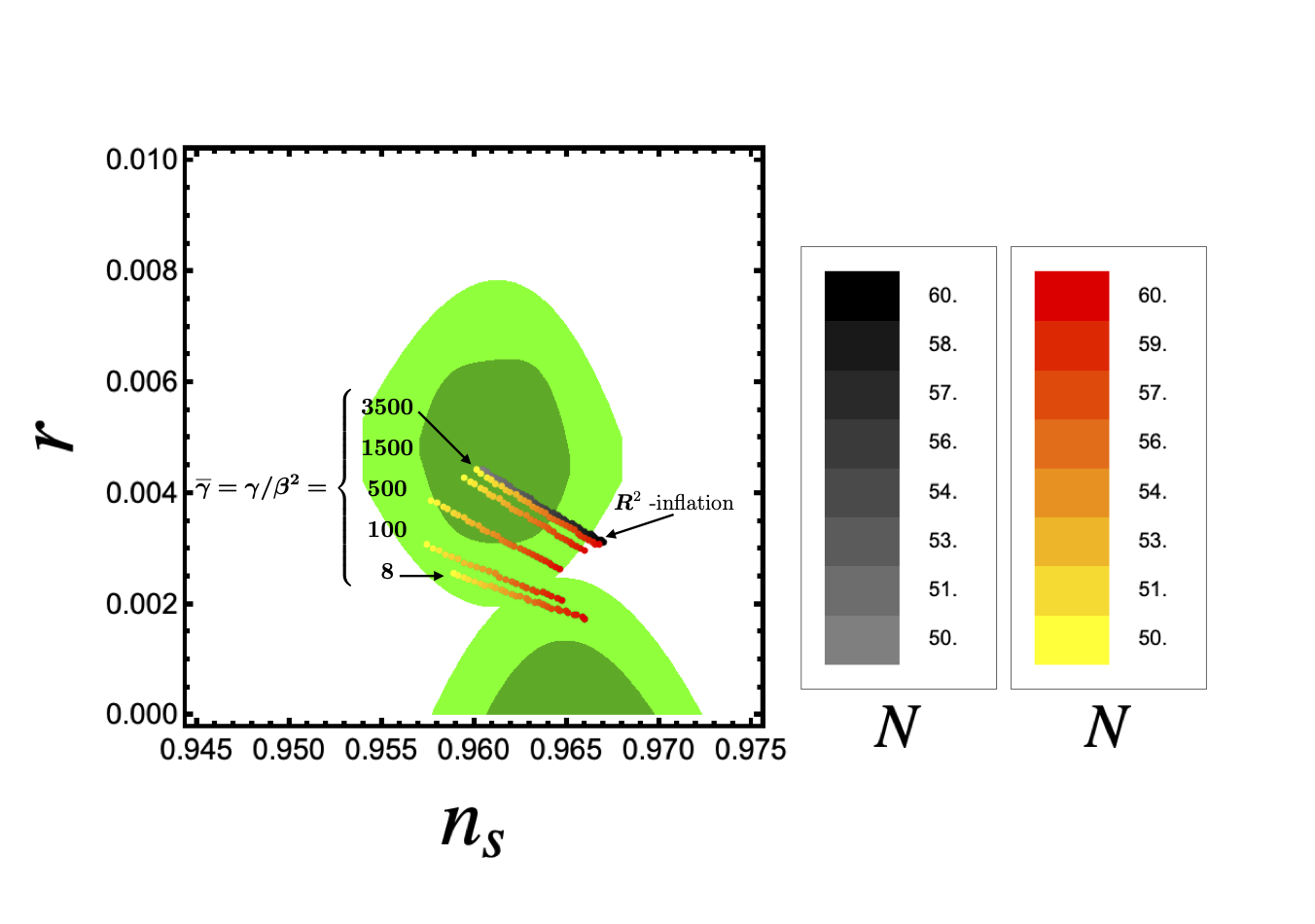}
\caption{The prediction (red-yellow points) in the $n_s$-$r$ plane,
where we have used $\bar{\gamma} $
$=3500, 1500, 500, 100$ and $8$  from top to bottom,, and
the color represents $N \in [50,60]$. The black-grey points are for
the Starobinsky inflation.
The green backgrounds are the LiteBird/Planck constraints \cite{LiteBIRD:2022cnt}
at $95$\% (light green) and $68$\% (green) confidence level;
the upper one shows the constraint
assuming the Starobinsky inflation with $N = 5 1$ as the fiducial model, while
the lower one shows the constraint
 in the case of null detection of $r$. }
\label{ns-r-N}
\end{center}
\end{figure}

We now proceed with the presentation of our results.
 Fig.~\ref{ns-r-N} shows the prediction (red-yellow points) in the $n_s$-$r$ plane for
$\bar{\gamma} =3500\,,1500,500,100,8$  (from top to bottom),
where the color represents $N$. 
\footnote{The lower bound for $N$ coming from
$T_\text{RH} \gtrsim 2\times 10^9$ GeV for a viable
leptogenesis  with $m_N  \gtrsim 2\times 10^7$ GeV \cite{Giudice:2003jh} can not be applied to the present model,
because the scalar $S$ (which is a part of the inflaton) 
 couples to the right-handed neutrinos and can directly reheat them.} The black-grey points are for
the Starobinsky inflation.
It is interesting to observe that the tensor-to scalar ratio $r$ of the present model
is smaller than that of the Starobinsky inflation.
The green backgrounds are the LiteBird/Planck constraints \cite{LiteBIRD:2022cnt}
at $95$\% (light green) and $68$\% (green) confidence level;
the upper one shows the constraint
assuming the Starobinsky  inflation with $N = 5 1$ as the fiducial model, while
the lower one shows the constraint
 in the case of null detection of $r$. 
As we see from Fig.~\ref{ns-r-N}, there exists a parameter space of our model
which will not be excluded
by LiteBird/Planck in  the case of null detection.
We will arrive at the same conclusion even if we will 
include the CMB-S4 constraint \cite{CMB-S4:2020lpa,Achucarro:2022qrl}.

Of the scale invariant models cited in  
\cite{Salvio:2014soa}-\cite{Kubo:2022dlx}, the models of Refs. 
\cite{Salvio:2014soa, Salvio:2020axm, Kannike:2015apa,Farzinnia:2015fka,Karam:2018mft,Kubo:2020fdd,
Ferreira_2019,Tambalo_2017,Barnaveli:2018dxo,Ghilencea:2019rqj,Kubo:2018kho,
Vicentini:2019etr,Gialamas:2020snr,Gialamas:2021enw,Aoki:2021skm,Kubo:2022dlx} can be regarded as  an extension of 
 the  Starobinsky model and in fact predict small values of $r$.
 However, only those of Refs.
 \cite{Salvio:2020axm,Tambalo_2017,Ghilencea:2019rqj,Gialamas:2020snr,
 Gialamas:2021enw,Aoki:2021skm} 
 predict $r$ which is smaller than that of the  Starobinsky model.
 Therefore, the null observation at LiteBird and CMB-S4 can exclude 
 many models based on an scale invariant extension 
 of the Starobinsky model.

\begin{figure}[ht]
\begin{center}
\includegraphics[width=14cm]{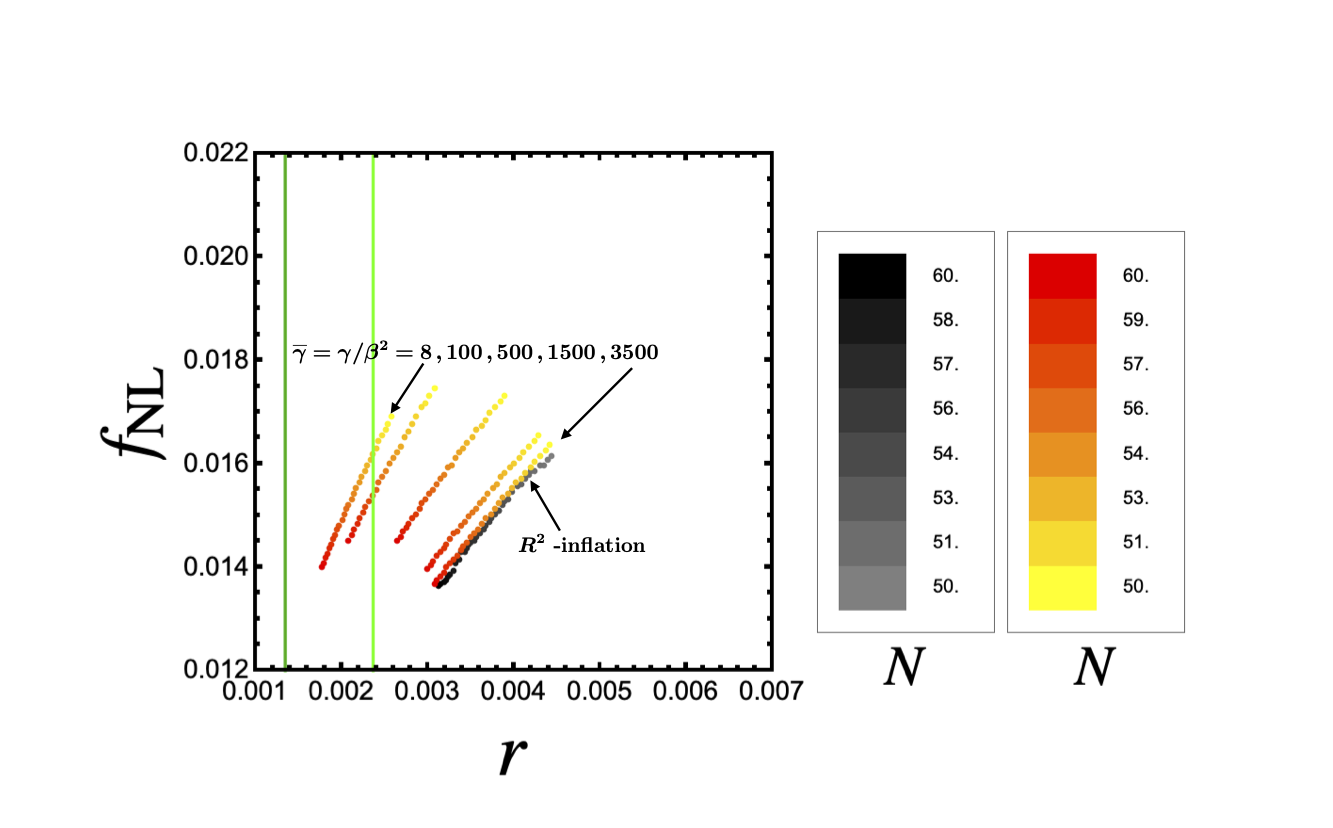}
\caption{The prediction in the $r$-$f_{NL}$ plane
 for, from right to left, $\bar{\gamma} =3500\,,1500,500,100,8$, where
the black-grey points are for  the Starobinsky inflation. 
The green (dark green) vertical line
is the upper bound of $r$ at $95 \,(68) $\% confidence level
in the case of null detection \cite{LiteBIRD:2022cnt}.}
\label{fNL-r-ns}
\end{center}
\end{figure}

Fig.~\ref{fNL-r-ns} shows the prediction in the $r$-$f_{NL}$ plane
 for
$\bar{\gamma} =3500\,,1500,500,100,8$ (from right  to left), where
the black-grey points are for the Starobinsky inflation. 
The green (dark green) vertical line
is the upper bound of $r$ at $95 \,(68) $\% confidence level
in the case of null detection \cite{LiteBIRD:2022cnt}.
As we see from Fig. \ref{fNL-r-ns}, the non-Gaussianity parameter $f_{NL}$
is $O(10^{-2})$, so that it will be  too small to be observed \cite{Komatsu:2001rj,Achucarro:2022qrl}.

\begin{figure}[ht]
\begin{center}
\includegraphics[width=11cm]{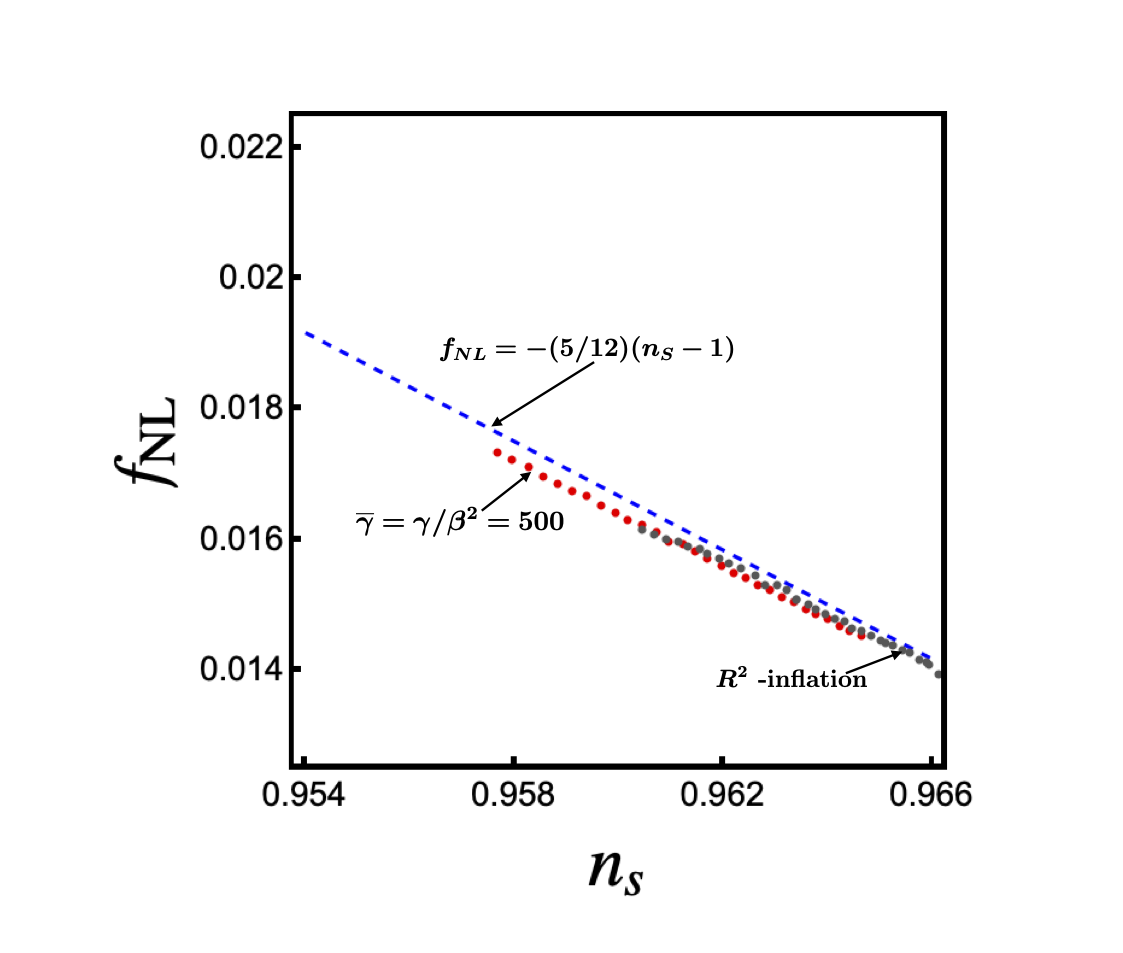}
\caption{$f_{NL}$ as a function $n_s$ for $\bar{\gamma}=500$ (red), where
the grey points are for the Starobinsky inflation.
The blue dotted line stands for  single-field inflation \cite{Maldacena:2002vr,Creminelli:2004yq}.}
\label{fNL-ns}
\end{center}
\end{figure}

Fig.~\ref{fNL-ns}  shows $f_{NL}$ as a function of
$n_s$ for $\bar{\gamma}=500$, where
the gray  points are  for the Starobinsky inflation.
In the case of single-field inflation we have $f_{NL}=-(5/12) (n_s-1)$ \cite{Maldacena:2002vr,Creminelli:2004yq}
which is presented by the blue dotted line.
As we have expected and we see from Fig.~\ref{fNL-r-ns}, our multi-field inflation model effectively behaves
as a single-filed model.

\section{Dark matter}
\label{darkmatter}

Because of the vector-like flavor symmetry
(i.e., $SU(3)_V$ or its subgroup), 
the CP-odd scalars $\phi_a$ (the quasi-NG bosons) are stable and hence
are good a DM candidate. 
Dark mater can be produced during or after the reheating phase (see e.g.,
Refs.~\cite{Chung:1998rq,Allahverdi:2002nb,Garcia:2020eof}).
It has been found in Ref.~\cite{Aoki:2021skm}
that $m_\phi > m_S$ is satisfied in the most of the parameter space,
so that the inflaton $S$ can not decay into these scalars.
However, if we  break $SU(3)_V$ down to $SU(2)_V\times U(1)$
 and  assume a
 hierarchy in the Yukawa couplings   \cite{Ametani:2015jla}, i.e.,
 \al{
 {\boldsymbol y} &= \mbox{diag.}\,(y_1\,,\,y_1\,,\,y_3\,)~
 \mbox{with}~ y_1=y_2< y_3\,, \label{hierarchy}
 }
 where  ${\boldsymbol y}$ is the Yukawa matrix in
 the hidden sector described by the Lagrangian (\ref{LH}),
 there is a sufficiently large parameter space in which 
 $S$ can  decay into the lightest meson which we will identify as DM.
 To see this, we note that under (\ref{hierarchy}) 
 the quasi-NG bosons fall into three categories, $\tilde{\pi}=\left\{ \tilde\pi^\pm,~\tilde{ \pi}^0 \right\}, 
\tilde{K} =\left\{\tilde{K}^\pm,~\tilde{K}^0, ~\bar{\tilde{K}}^0 \right\}$
and $\tilde{\eta}$, where
 \begin{align}
\tilde{\pi}^\pm &\equiv (\phi_1\mp i\phi_2)/\sqrt{2}~,~~
\tilde{ \pi}^0 \equiv \phi_3~, ~\nn\\
 \tilde{K}^\pm & \equiv (\phi_4\mp i  \phi_5)/\sqrt{2}~,
 ~~\tilde{K}^0(\bar{\tilde{K}}^0) \equiv
  (\phi_6+(-) i \phi_7)/\sqrt{2}~,~~ \tilde{\eta}^8 \equiv \phi_8\,,
  \label{mesons}
  \end{align}
and $\tilde{\eta}^8$ will mix with $\tilde{\eta}^0=\phi_0
$ to form the mass eigenstates
$\tilde{\eta}$ and $\tilde{\eta}'$. 
The states in the same category have the same mass, 
$m_{\tilde{\pi}^0}=m_{\tilde{\pi}^\pm} ( \equiv  m_{\tilde{\pi}}) $ and 
$m_{\tilde{K}^\pm}=m_{\tilde{K}^0}=m_{\bar{\tilde{K}}^0} ( \equiv m_{\tilde{K}} )$,
with $m_{\tilde{\pi}} < m_{\tilde{K}} < m_{\tilde{\eta}}$. In the following discussion
we will work in the parameter space, in which 
\be
m_\text{DM}\equiv m_{\tilde{\pi}} \ll m_S < m_{\tilde{K}} < m_{\tilde{\eta}}
\label{DM-mass}
\ee
is realized and use the approximation for $m_\text{DM}
 (= m_{\tilde{\pi}})$  \cite{PhysRev.175.2195}, 
 $m^2_\text{DM }/m_{\phi}^2 \simeq m_u/m_q\simeq y_1/y$,
 where $m_u$ is the current quark mass in the $SU(2)_V\times U(1)$ case, and
$m_q$, $m_{\phi}$ and $y$ are the current quark mass,
the dark meson mass and the Yukawa coupling, respectively, in the $SU(3)_V$ limit.

In the parameter space, in which  the mass hierarchy (\ref{hierarchy})
is satisfied, the inflaton $S$
can decay only into a pair of DM particles - but not into the other mesons,
and therefore the heavier dark mesons are
 not produced during the reheating stage and later~\cite{Allahverdi:2002nb}.
 The decay width  for $S\to \tilde{\pi}+ \tilde{\pi}$ is given by
\al{
\gamma_\text{DM}&=
\frac{3 \,G_{\tilde{\pi}\tilde{\pi} S}^2}{16\pi m_S} 
\sqrt{1-\frac{4 m_\text{DM}^2} {m_S^2}  }\,,
\label{gammapi}
}
where the effective coupling  $G_{\tilde{\pi}\tilde{\pi} S}$ is calculated  in Ref.~\cite{Ametani:2015jla} and is found to be
$G_{\tilde{\pi}\tilde{\pi}S}/\Lambda_H
\simeq -0.012\,y_1$ for $y_3=0.0046$ and $y_1 \ll  y_3$.
The DM can also be produced by the co-annihilation of $N$, i.e.,
$N\, N\leftrightarrow \tilde{\pi}\,\tilde{\pi} $. However,  its cross section is
 small due to the small $y_M$ (see the subsection \ref{four-elements}).  
We have found that the DM production rate by
the co-annihilation process is negligibly small
compared with that through the decay of $S$.

Under the situation specified above 
 we arrive at a system, which consists of 
only the inflaton $S$ and the dark matter  $\tilde{\pi}$ and in which 
the evolution of their number densities,  $n_S$ and $n_\text{DM}$, can be described
by the coupled Boltzmann equations~\cite{Chung:1998rq}
\al{
\frac{d n_S}{dt} &= -3 H n_S-\Gamma_S\, n_S \,,
\label{nS}\\
\frac{d n_\text{DM}}{dt}&= -3 H n_\text{DM}
+\gamma_\text{DM} \,n_S 
\label{npi}
}
with
$\Gamma_S$ being the total decay width of $S$.
Eq.~(\ref{nS})  can be simply solved \cite{Kolb:1990vq}:
\al{
n_S (a) &= \frac{\rho_\mathrm{end}}{m_S}\,\left[
\frac{a_\mathrm{end}}{a}\right]^3\, e^{-\Gamma_S\,(t-t_\mathrm{end})}\,,
\label{nSa}
}
where $a$ is the scale factor at $t > t_\mathrm{end}$, $a_\mathrm{end}$
is $a$ at the end of inflation $t_\mathrm{end}$ and 
$\rho_\mathrm{end}=\rho_S(a_\mathrm{end})=m_S n_S(a_\mathrm{end})$
is the inflaton energy density at $t_\mathrm{end}$.
Then we follow Refs.~\cite{Martin:2010kz,Lozanov:2017hjm,Planck:2018jri}
to
proceed with the  calculation of  the DM relic abundance
$\Omega_\text{DM} h^2$ and define
%
the reheating temperature $T_\mathrm{RH}$ as
\al{
\rho_\mathrm{RH}
&= \frac{\pi^2}{30}\,g_\mathrm{RH} \,T_\mathrm{RH}^4\,,
\label{TRH}
}
where  $g_\mathrm{RH}$ is  the relativistic degrees of 
  freedom at the end of reheating. Finally, we arrive at
~\cite{Allahverdi:2002nb,Aoki:2021skm}:
\al{
\Omega_\text{DM} h^2 
&\simeq 2.04\times 10^{8}\,
\left(\frac{\gamma_\text{DM}}{\Gamma_S}\right) \left(\frac{m_\text{DM}}{m_S}\right) \,\frac{T_\mathrm{RH}}{1\,\mbox{GeV}}\,,
\label{OmegaDM}
}
where the branching ratio $\gamma_\text{DM}/\Gamma_S$ can be obtained
from $\gamma_\text{DM}$ 
given in Eq.~(\ref{gammapi}) together with the assumption that
$1/\Gamma_S$ is the time scale at the end of the reheating phase~\cite{Kolb:1990vq,Chung:1998rq}, which means\ $1/H(a_\mathrm{RH})
=\left(\,3\, M_\mathrm{Pl}^2/\rho_\mathrm{RH}
\,\right)^{1/2}$. 

\begin{figure}[ht]
\begin{center}
\includegraphics[width=12cm]{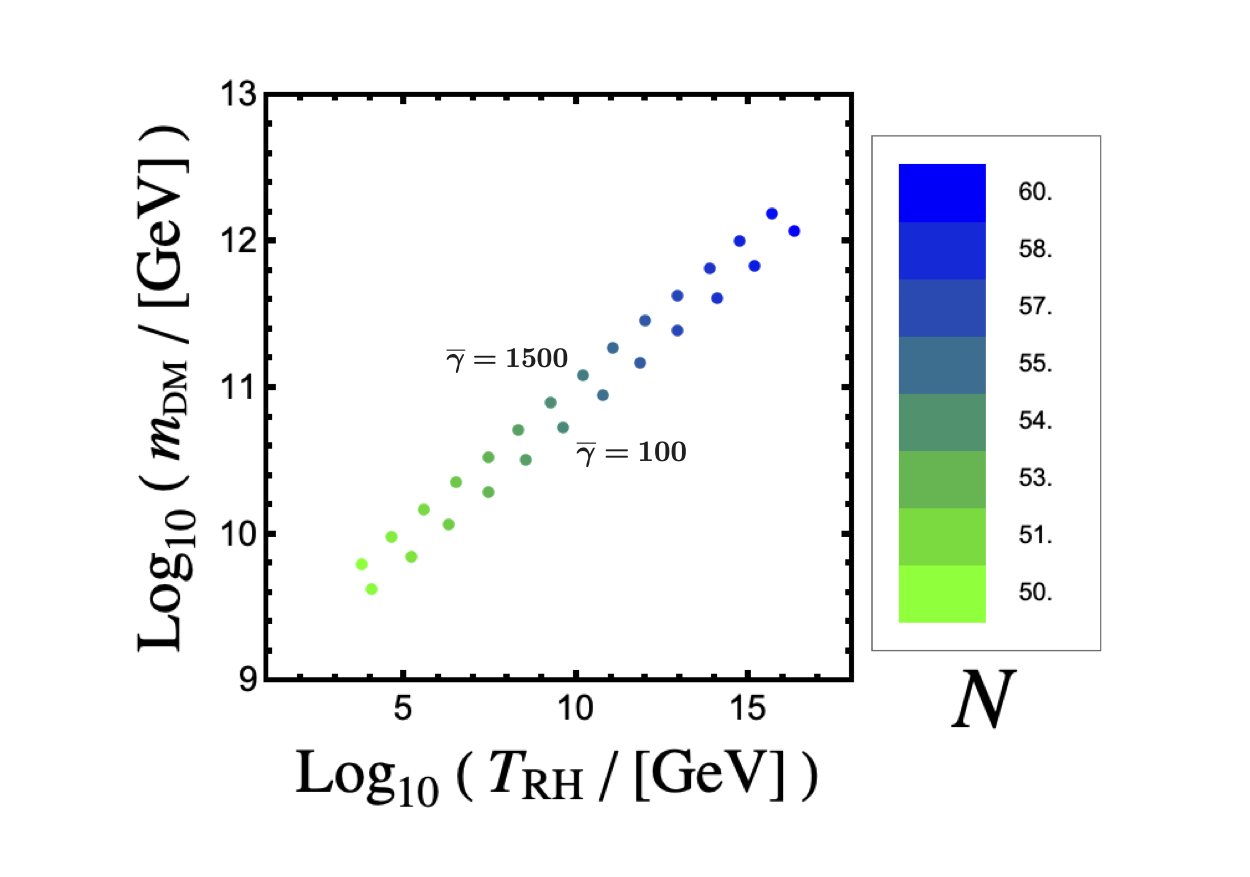}
	\caption{Dark matter mass $m_\text{DM}$ against the 
		reheating temperature $T_\mathrm{RH}$
		for $\bar{\gamma}=1500$ and $100$ with
		$y_3=0.0046$ and $\lambda_S=0.0114$.
		The colored points satisfy  
		 $\Omega_\text{DM} h^2=0.1198\pm 0.0024\,(2\sigma)$,
		 where the color represents the e-foldings $N$.		
		 The corresponding $r$ and $n_s$ can be found in Fig. \ref{ns-r-N}.}

\label{DM-TRH}
\end{center}
\end{figure}

Although
 $\rho_\mathrm{RH}$ and hence $T_\mathrm{RH}$
 are unknown quantities,
it is possible to constrain
$T_\text{RH}$ for a given inflation model without specifying  reheating 
mechanism~\footnote{ The inflaton $S$ can reheat the SM sector 
through the Yukawa coupling
$S$-$N$-$N$, 
 where the size of the coupling 
 is very small in the present model, i.e., $y_M\sim 10^{-9} $,
because we implement the neutrino option mechanism to break the SM gauge symmetry. Another interesting way to reheat the SM sector 
 may follow
from the observation that the kinetic terms become non-canonical when going
from the Jordan frame to the Einstein, such that the the inflaton,
especially the scalaron $\varphi$ in our model, interacts with the SM in this frame
\cite{Csaki:2014bua,Kannike:2015apa}.
}  \cite{Liddle:2003as,Martin:2010kz,Martin:2013tda,Planck:2018jri}:
\begin{align}
N &=\ln \left(\frac{a_\mathrm{end}}{a_*}\right)\nonumber\\
&=
66.89-\frac{1}{12}\ln g_\mathrm{RH}+\frac{1}{12}\ln\left(\frac{\rho_\text{RH}}{\rho_\text{end}}\right)+\frac{1}{4} \ln \left(\frac{V_*^2}{M_\mathrm{Pl}^4 \,\rho_\mathrm{end}}  \right) -\ln \left(\frac{k_*}{a_0 H_0}  \right)\,,
\label{Ne}
\end{align}
where $a_*=k_*/H_*$ is the scale factor at the time of CMB horizon 
exit $(t=t_*)$,$V_*=V(t=t_*)$,
$k_*$ is the pivot scale set by the Planck
mission \cite{Aghanim:2018eyx,Planck:2018jri}, 
and $H_*$ is the Hubble parameter at $a=a_*$.
Further,  we use
\begin{equation}
\rho_\mathrm{end} =
\frac{V_\mathrm{end}(3-\varepsilon_*)}{(3-\varepsilon_\mathrm{end})}\,,
\end{equation}
where
$V_\mathrm{end}=V(t=t_\mathrm{end})$,
$\varepsilon_\mathrm{end}=\varepsilon(t=t_\mathrm{end})$, 
$\varepsilon_*=\varepsilon(t=t_*)$
in Eq.~(\ref{Ne}).~
\footnote{The energy scale (temperature)  just after the end of inflation 
 is approximately $E_\text{max}\simeq  \rho^{1/8}_\text{end}
\,(\sqrt{8\pi}M_\text{Pl}
\Gamma_S/g_\text{RH})^{1/4}$ ($g_\text{RH}\simeq 10^2$) \cite{Chung:1998rq}.
We find  $E_\text{max}/\Lambda_H \simeq 4.1\times 10^{-5} <1$
for the benchmark point (\ref{bench4}), which means  that the applicability of 
the NJL theory as an effective theory is not violated,
according to the discussion in \ref{NJL}.
We have made this consistency check for other sets of parameters
and found that the above requirement is satisfied for all the cases.}
In Fig.~\ref{DM-TRH}  we show the points in 
the $T_\text{RH}-m_\text{DM}$ plane,  
for which $\Omega_\text{DM} h^2=0.1198\pm 0.0024\,(2\sigma)$
is obtained. 
The dark matter mass $m_\text{DM}$ is calculated from
$m^2_\text{DM}\simeq (y_1/y_3) m_\phi^2$, where $m_\phi$ is the meson mass
in the $SU(3)_V$ limit (with $y=y_3$) and obtained in the NJL formalism \cite{Aoki:2021skm}.
Since  $\Omega_\text{DM}h^2$  (\ref{OmegaDM})  depends on 
$m_\text{DM}$ and $\gamma_\text{DM}$ and hence
on $y_1$ and is strongly constrained,
 they are closely correlated.
We have varied   $y_1$ 
for  $\bar{\gamma}=\gamma/\beta^2=100\,\mbox{and}\,1500$ 
with $\lambda_S$ and $y_3$ fixed at $0.0114$ and $0.0046$, respectively.
(The corresponding $r$ and $n_s$ can be found in Fig.~\ref{ns-r-N}.)

As we see from Fig.~\ref{DM-TRH}, our DM is heavier than $\sim 10^9$ GeV.
Furthermore, there exist only indirect couplings with the SM sector:
In addition to the gravitational interaction, the scalar $S$ mediates
an interaction,
as we can see from the Lagrangians (\ref{LSM}) and (\ref{LNR}).
But the coupling of $S$ with the SM is extremely suppressed;
the Higgs portal coupling $\lambda_{HS}\sim O(10^{-28})$
and the Yukawa coupling  $y_M\sim O(10^{-9})$, implying
that ourDM will be invisible, 
except if  it can decay into SM particles  \cite{Das:2023wtk}, such that
the decay is consistent with, e.g., the resent observation
of an extremely energetic cosmic ray of $O(10^{11})$ GeV at Telescope Array
\cite{TelescopeArray:2023sbd}.

\section{Conclusion}
\label{conclusion}
In this paper we have assumed that the origin of
the Planck mass and the electroweak scale
 including the right-handed neutrino mass is 
the  chiral symmetry breaking in a QCD-like hidden sector which
couples with  the  SM sector only via a real scalar
$S$, the mediator.
The scalar $S$ is not only a mediator, but also an inflaton,
which makes a Higgs-inflation-like scenario possible.

The inflationary system that we have considered contains three scalar fields;
$\sigma$ (the chiral condensate in the NJL formalism),
 $S$ (the mediator) and  $\varphi$ (the scalaron).
The scalar potential (\ref{VSphi}) that depends on these fields
has a river-valley like structure, and 
we have found that $\sigma$  changes only slightly along the  river,
so that we have assumed that it stays during inflation at the position
of the  absolute minimum of the potential.
In this way we have reduced the system to a two-field system
of $S$ and $\varphi$.
Thanks to the  river-valley like structure of the potential,
the trajectories in the two-field system 
converge very fast to a fixed-point  trajectory. If they converge to a fixed-point  trajectory  before the horizon exit, the dependence of the initial field values
has only negligible effects on  the observable inflationary parameters -
a nice feature of  river-valley like potential.

We have analyzed the  river-valley like structure of the two-field system 
in detail  to separate the heavy and light modes, because 
one of the slow-roll conditions
is violated in the heavy mode direction of the potential
and only the light mode contributes to the inflationary parameters.
This implies that the system is effectively a single-field
system, as we have expected from Ref. \cite{Aoki:2021skm}.
Nevertheless we have applied the $\delta N$  formalism
\cite{Sasaki:1995aw,Nakamura:1996da,Sasaki:1998ug,Lyth:2004gb,Lyth:2005fi,Sugiyama:2012tj}
to compute the inflationary parameters, including 
the non-Gassianity parameter $f_{NL}$  of the primordial curvature perturbations.
To compute the inflationary parameters of our model concretely
we have  adjusted   the $\delta N$ formalism to the river-valley like structure of the 
potential and developed an algorithm 
 to compute  numerically the derivatives of
$N$ with respect to background fields.

Though
the scalar spectral index $n_s$ and the tensor-to-scalar ratio $r$
of the present model
are similar to those
of  the Starobinsky inflation \cite{Starobinsky:1980te,Mukhanov:1981xt,Starobinsky:1983zz},
we have found that 
$r$ is smaller than that of the Starobinsky inflation and
 found that 
the present model will be consistent with a null detection of  $r$
at LiteBird/CMB-S4, while the Starobinsky inflation will  be excluded
in this case  \cite{LiteBIRD:2022cnt}.

Since we have effectively a single-field system at hand, 
the non-Gaussianity parameter $f_{NL}$ is expected 
to be $O(10^{-2})$ \cite{Maldacena:2002vr,Creminelli:2004yq},
which we have explicitly confirmed by using the $\delta N$ formalism.
This means that the Universe described by the multi-field cosmological system
of the model in this paper
is so  isotropic, that 
the non-Gaussianity will not be measured in future experiments
\cite{Komatsu:2001rj,Achucarro:2022qrl}.
It should be, however, noted that a considerable improvement 
of the experimental accuracy of $f_{NL}$ as well as of $r$
can be achieved 
by using  the  fluctuations in the 21-cm signal from atomic hydrogen 
in the dark ages  \cite{,Munoz:2015eqa,Book:2011dz}.

NG bosons are produced
due to the  chiral symmetry breaking in the hidden sector
 in a very much similar way as in QCD. 
Since the Yukawa coupling of $S$
with the hidden  fermions explicitly breaks the chiral symmetry, they are
 quasi-NG bosons and massive. They
 are stable because of the unbroken vector-like flavor symmetry and 
 therefore can be a DM candidate.
We have however 
realized that the full $SU(3)_V$  flavor group has to be explicitly broken
in order to make 
the only viable scenario for a realistic DM  in our model, 
the decay of the inflaton $S$ into them, possible. 
The dark matter particles have turned out be  heavier than $\sim 10^9$ GeV,
and  there exist only indirect couplings with the SM sector
in addition to the gravitational interaction.
Unfortunately, the indirect coupling with the SM,
which is mediated by $S$,  is so suppressed,
that the dark matter particles will be  unobservable
in direct as well as indirect measurements.

\section*{Acknowledgments}
J.~Y. thanks Yi Wang for helpful discussions on the $\delta N$ formalism.
The work is supported in part by JSPS KAKENHI
Grant No. 20H00160 and 23K03384 (M.~A.) and
Grant No. 23K03383 (J.~K.).

\bibliographystyle{JHEP}
\bibliography{uosref1}
\end{document}